\documentclass[pre, aps, twocolumn, a4paper, superscriptaddress]{revtex4-1}

\usepackage{amsmath}
\usepackage{amssymb}
\usepackage{amsthm}

\usepackage{graphicx}
\usepackage{enumitem}   
\usepackage[english]{babel}
\usepackage{float}

\usepackage{nicematrix}
\usepackage{makecell}
\usepackage{xcolor}
\usepackage{hyperref}
\usepackage{tikz}
\usepackage{mhchem}
\usepackage{array}

\usepackage{libertine}

\usepackage{easyReview}

\usepackage{bbm}

\usepackage{tabularray}
\usepackage{layouts}

\newcommand{\E}{\mathcal{E}}
\newcommand{\Ef}{\mathcal{E}_f}
\newcommand{\Ep}{\mathcal{E}_p}

\newcommand{\Z}{\mathcal{Z}}
\newcommand{\Y}{\mathcal{Y}}
\newcommand{\X}{\mathcal{X}}

\newcommand{\Zp}{\mathcal{Z}_p}
\newcommand{\Yp}{\mathcal{Y}_p}
\newcommand{\Xp}{\mathcal{X}_p}

\newcommand{\Zf}{\mathcal{Z}_f}
\newcommand{\Yf}{\mathcal{Y}_f}
\newcommand{\Xf}{\mathcal{X}_f}

\newcommand{\Nabla}{\boldsymbol{\nabla}}
\newcommand{\NablaER}{\boldsymbol{\nabla}^\E_{\R}}
\newcommand{\NablaEnR}{\boldsymbol{\nabla}^\E_{\nR}}

\renewcommand{\S}{ \mathbb{S} }
\newcommand{\invS}{\mathbb{G}}
\newcommand{\Sbar}{\widehat{\mathbb{S}}}
\newcommand{\T}{\mathbb{T}}

\newcommand{\R}{\mathcal{R}}
\newcommand{\nR}{\backslash\mathcal{R}}

\newcommand{\Mu}{\boldsymbol{\mu}}
\newcommand{\I}{\boldsymbol{I}}
\newcommand{\J}{\mathcal{J}}

\renewcommand{\j}{\boldsymbol{j}}

\renewcommand{\L}{\mathbb{L}}
\newcommand{\Lbar}{\widehat{\mathbb{L}}}

\newcommand{\M}{\mathbb{M}}

\renewcommand{\d}{\mathrm{d}}

\newcommand{\e}{\boldsymbol{e}}
\newcommand{\z}{\boldsymbol{z}}
\newcommand{\x}{\boldsymbol{x}}
\newcommand{\y}{\boldsymbol{y}}

\newcommand{\G}{\mathcal{G}}
\newcommand{\F}{\boldsymbol{\mathcal{F}}}

\newcommand{\wprod}{\dot{W}_\text{prod}}

\DeclareMathAlphabet{\mathsfit}{T1}{\sfdefault}{\mddefault}{\sldefault}
\SetMathAlphabet{\mathsfit}{bold}{T1}{\sfdefault}{\bfdefault}{\sldefault}

\DeclareTextFontCommand{\sf}{\sffamily\itshape}

\graphicspath{{./Figures/}}

\begin{document}
	
\title{
	Nonequilibrium properties of autocatalytic networks
}

\author{Armand Despons}
\affiliation{Gulliver Laboratory, UMR CNRS 7083, PSL Research University, ESPCI, Paris F-75231, France}

\date{\today}

\begin{abstract}
	Autocatalysis, the ability of a chemical system to make more of itself, is a crucial feature in metabolism and is speculated to have played a decisive role in the origin of life.
	Nevertheless, how autocatalytic systems behave far from equilibrium remains unexplored.
	In this work, we elaborate on recent advances regarding the stoichiometric characterization of autocatalytic networks, particularly their absence of mass-like conservation laws, to study how this topological feature influences their nonequilibrium behavior.
	Building upon the peculiar topology of autocatalytic networks, we derive a decomposition of the chemical fluxes, which highlights the existence of productive modes in their dynamics.
	These modes produce the autocatalysts in net excess and require the presence of external fuel/waste species to operate.
	Relying solely on topology, the fluxes decomposition holds under broad conditions and, in particular, do not require steady-state or elementary reactions. 
	Additionally, we show that once externally controlled, the non-conservative forces brought by the external species do not act on these productive modes.
	This must be considered when one is interested in the thermodynamics of open autocatalytic networks. 
	Specifically, we show that an additional term must be added to the semigrand free-energy.
	Finally, from the thermodynamical potential, we derive the thermodynamical cost associated with the production of autocatalysts.
\end{abstract}


\maketitle



\section*{Introduction}

\subsection*{Context and motivations}

At the heart of cellular metabolism is its capability to replicate essential biomolecules from pre-existing ones, employing elementary compounds as elemental building materials. 
This trait emphasizes the vital role of autocatalysis, namely, systems that produce more of themselves, in living organisms.
It results in highly complex behavior \cite{Schuster2019}, enabling growth \cite{Roy2021, Lin2020, Sakref2024} and self-reproduction \cite{Ameta2021}.
These attributes are believed to have played a central role in abiogenesis \cite{Peng2022, Hordijk2014, Vincent2021, Peng2020, Xavier2020, Higgs2015, Bernhardt2012, Pavlinova2023}. 
The growing interest in autocatalytic networks can be traced back to the pioneering studies of S. Kauffman on autocatalytic sets \cite{Kauffman1971, Kauffman1986}. 
Since then, recent studies have elucidated the topological features of autocatalytic networks, allowing classification and detection in large networks \cite{Blokhuis2020, Andersen2021}.  

Concurrently, the thermodynamics of chemical systems has undergone intense investigations since the foundational works of Gibbs, who introduced the concept of chemical potential \cite{Gibbs1928}, and De Donder, who introduced the concept of affinity \cite{deDonder1927}, which quantifies the irreversible forces driving chemical systems. 
Specifically, in the 1970s, particular attention was drawn to the stochastic dynamics of chemical systems based on the chemical master equation \cite{McQuarrie1967, Gillespie1992}. 
This has allowed for the derivation of the foundational concepts regarding nonequilibrium thermodynamics of chemical systems \cite{Schnakenberg1976, Mou1986, Hill1983}.
Since then, the emergence and growing popularity of stochastic thermodynamics, combined with chemical reaction networks formalism, have led to the development of a consistent theory assessing the nonequilibrium thermodynamics of chemical networks \cite{Rao2016, Ge2016-2, Avanzini2021}. 
As living systems are arguably the most important example of nonequilibrium systems, this theory was quickly applied to study biochemical networks \cite{Alberty2003, Qian2005}.

More recently, there has been an increasing focus on establishing connections between topology and nonequilibrium dynamics in chemical networks \cite{Cengio2022, Harunari2022, Despons2023}. 
Indeed, since topology precedes dynamics, it imposes universal constraints on the network's behavior, which hold under a wide range of conditions \cite{Aslyamov2024, Owen2023}.
Yet, the connections between the specific topology of autocatalytic networks and their nonequilibrium behavior have not yet been explored. 
Therefore, this paper aims to investigate the interplay between the topology of autocatalytic networks and their nonequilibrium dynamics.

\subsection*{Outline of the paper}

To derive our results, we build upon the stoichiometric condition of autocatalysis given by Blokhuis \emph{et al.} in Ref.~\cite{Blokhuis2020}, recalled in Eq.~\eqref{eq:stoichiometric_condition}, and we introduce the notion of autocatalytic networks.  
Specifically, we will consider chemical systems where an autocatalytic network is coupled with external species acting as fuel and waste materials, serving to produce autocatalysts in excess and allow for mass-conservation in the system. 
These assumptions are captured by the topology such networks, encoded in their stoichiometric matrices, which follow the block decomposition given in Eq.~\eqref{eq:Nabla}. 
 
Then, in Section~\ref{Geometry}, by relying solely on the topology of such networks, we derive their nonequilibrium behavior. 
The main result of this section is the decomposition of the chemical fluxes in Eq.~\eqref{eq:decomposition_fluxes} on the unique linear basis distinguishing pathways that produce the autocatalysts in the autocatalytic network (productive modes) and pathways preserving the state in the autocatalytic network (cycles). 
As the analysis is solely based on stoichiometry, this decomposition is independent of the chemical fluxes. 
In addition, we show that the physical role of the chemical species exchanged with the environment is fully captured by the topology and depends whether or not they break a conservation law. 
Particularly, we demonstrate that, even though both the productive modes and the cycles of the autocatalytic network require influx of external species, they are in fact two drastically different processes. 
We will illustrate the main ideas of this section on the network associated with glucose metabolism and its regulation (glycolysis and gluconeogenesis).

In Section~\ref{Thermo}, we dwell on the thermodynamics of autocatalytic networks. 
Specifically, in this section we derive the nonequilibrium thermodynamical potential of autocatalytic networks.
It is obtained from the semigrand Gibbs free-energy by adding an extra term similar to the potential energy associated with conservative forces in Newtonian mechanics. 
The latter represents the conservative forces that the fuel and waste species exert on the reactions when they are converted into an excess of autocatalysts; it is expressed using a gauge transform of the chemical potentials. 
The main results of this section are then the expression of the thermodynamical potential in Eq.~\eqref{eq:G_def} and its associated EPR decomposition in Eq.~\eqref{eq:second_law_dt_G}. 
Finally, we focus on autonomous networks where the cost of sustaining the productive modes and the cycles takes a simple form.

\subsection*{Setup and notations}

\subsubsection*{Autocatalytic CRNs}

We consider an autocatalytic network composed of reacting chemical species that change through chemical reactions. 
We refer to the chemical species in the autocatalytic network as the \emph{autocatalytic species}, and we let $\Z$ be the set containing all the autocatalytic species. 
Similarly, the reactions in the autocatalytic network are called the \emph{autocatalytic reactions} that we gather in the set $\R$. 
The net change of species $z \in \Z$ along reaction $\rho \in \R$ is $S^z_\rho$: it is positive (resp. negative) if $z$ is produced in net excess (resp. consumed) by the reaction. 
Subsequently, the stoichiometric matrix $\S = \left\lbrace S^z_\rho \right\rbrace$ quantifies the net change of autocatalytic species (the rows of $\S$) along all the autocatalytic reactions (the columns of $\S$) \cite{Feinberg2019}. 
The autocatalytic nature of the network should be reflected in its topology, which is encoded in the stoichiometric matrix $\S$.
Specifically, following Blokhuis \emph{et al.} \cite{Blokhuis2020}, the stoichiometric condition for autocatalysis is the existence of a pathway creating a net excess of \emph{all} the autocatalytic species.
Such a pathway consists of a linear combination of the autocatalytic reaction, $\boldsymbol{u}$, which can be represented as an overall reaction: 
\begin{equation*}
	\ce{
		$\boldsymbol{\alpha} \cdot \z$ ->[\boldsymbol{u}] $\boldsymbol{\beta} \cdot \z $,
	}
\end{equation*}
in which $\z$ contains the autocatalytic species,  and where $\boldsymbol{\alpha}$ (resp. $\boldsymbol{\beta}$) stores the reactants (resp. products) stoichiometric coefficients. 
As a result, the stoichiometric condition prescribes that the net change of autocatalytic species along the overall reaction, \emph{i.e.}
$\S \cdot \boldsymbol{u} = \boldsymbol{\beta} - \boldsymbol{\alpha}$,   
contains only strictly positive components: 
\begin{equation}
	\label{eq:stoichiometric_condition}
	\S \cdot \boldsymbol{u} > 0,
\end{equation} 
component-wise. 

For simplicity, we will assume in the main text that the rows of $\S$ are linearly independent: 
\begin{equation}
	\label{eq:simpler_condtion}
	\ker ~ \S^\top = \left\lbrace \boldsymbol{0} \right\rbrace. 
\end{equation}
This stronger condition is sufficient (yet not necessary) to satisfy the stoichiometric condition in Eq.~\eqref{eq:stoichiometric_condition} and is observed for most autocatalytic networks.
We relax this assumption in the appendices to address the most general framework of Eq.~\eqref{eq:stoichiometric_condition} (Appendix A \& B).

\subsubsection*{External species}

The existence of a strictly productive reaction imposed by the stoichiometric condition Eq.~\eqref{eq:stoichiometric_condition} violates mass conservation. 
Hence, the autocatalytic network cannot exist independently and must be coupled to \emph{external species}, which also partake in the autocatalytic reactions as food/waste species or external catalysts.
In addition, the external species might also be involved in some \emph{additional reactions}.
In the following, we let $\E$ be the set of the external species. 
As a result, the topology of the network formed by considering both types of species (external and autocatalytic) is encoded in its stoichiometric matrix, $\Nabla$, which has the following block decomposition:
\begin{equation}
	\label{eq:Nabla}
	\boldsymbol{\nabla} = \quad 
	\begin{pNiceArray}{m{1em}m{1em}|m{1em}m{1em}}[last-row, first-col]
		\Vdots[line-style={solid, <->}, shorten=2pt]_{\small \rotatebox{90}{$\E$} }  & \Block{2-2}{\NablaER} &  & \Block{2-2}{\NablaEnR} & \\
		& & & & \\
		\Hline
		\Vdotsfor{3}[line-style={solid, <->}, shorten=2pt]_{\small \rotatebox{90}{$\Z$} } \hspace*{1em} & \Block{3-2}{\S} & & \Block{3-2}{ \boldsymbol{0} } & \\[-0.5em]
		&  &  &  &   \\
		& & & & \\
		& \Hdotsfor{2}[line-style={solid, <->}, shorten=1pt]_{ \R } &  \Hdotsfor{2}[line-style={solid, <->}, shorten=1pt]_{ \nR }  
	\end{pNiceArray}.
\end{equation}
In this decomposition, $\NablaER$ stands for the restriction of $\Nabla$ on the subsets of external species ($\E$) and autocatalytic reactions ($\R$), which quantifies how the external species participate in the autocatalytic reactions. 
In contrast, the additional reactions ($\nR$) involve only external species.
In practice, given a (possibly large) chemical network described by $\Nabla$, locating autocatalytic sub-network amounts to finding a subset $\Z$ of species and a subset $\R$ of reactions such that $\S \equiv \Nabla^\Z_\R$ follows the stoichiometric condition Eq.~\eqref{eq:stoichiometric_condition} (or its stronger counterpart Eq.~\eqref{eq:simpler_condtion}).

\subsubsection*{Dynamics}

We call $\j$ (resp. $\boldsymbol{v}$) the chemical fluxes associated with the autocatalytic reactions $\R$ (resp. additional reactions $\nR$).
The dynamics of the open CRN follow the kinetic rate equations \cite{Epstein1998} which, from the block decomposition of the stoichiometric matrix, can be written as:
\begin{gather}
	\label{eq:dyn_external}
	\d_t [\e]  ~=~ \NablaER \cdot \j ~+~ \NablaEnR \cdot \boldsymbol{v}  ~+~ \I^{\E}, \\[0.5em]
	\label{eq:dyn_autocat}
	\d_t [\z]  ~=~ \S \cdot \j ~+~ \I^{\Z},
\end{gather}
where $[\boldsymbol{s}]$, for $s \in S$ (with $S = \Z$ or $\E$), is the vector of chemical concentrations and $\I^S$ the external fluxes coupling the CRN with the environment. 
In what follows, we will assume that all the external species are exchanged with the environment, $I^e \neq 0$ for all $e \in \E$.
On the contrary, for the autocatalytic species, we let $\Y \subset \Z$ (resp. $\X \subset \Z$) be the subset of autocatalytic species subjected to a non-vanishing (resp. vanishing) external flux, \emph{i.e.} $I^y \neq 0$ for all $y \in \Y$ and $\I^\X = \boldsymbol{0}$. 
As a result, the dynamics in the autocatalytic sub-network is
\begin{equation}
	\label{eq:dynamics_subnetwork}
	\begin{gathered}
		\d_t [\y] = \S^{\, \Y} \cdot \j + \I^\Y, \\[.5em]
		\d_t [\x] = \S^{\, \X} \cdot \j.
	\end{gathered}
\end{equation}

\subsubsection*{Networks of elementary reactions}

In Section \ref{Thermo} we analyze the thermodynamics of the open sub-network. 
In doing so, for this section, we will further assume that the network is composed of elementary reactions.
In such networks, whenever a reaction exists, its backward counterpart must also exists from microscopic reversibility. 
Hence, the stoichiometric matrix and the elementary fluxes can be written as the result of a forward and backward contribution:
\begin{align}
	\label{eq:Nabla_pm}
	\Nabla = \Nabla_- - \Nabla_+, & &
	\j = \j^+ - \j^- , &  &
	\boldsymbol{v} = \boldsymbol{v}^+ - \boldsymbol{v}^- ,
\end{align}
where $\Nabla_-$ (resp. $\Nabla_+$) describes the products (resp. reactants) stoichiometry and $\boldsymbol{j}^+$/$\boldsymbol{v}^+$ (resp. $\boldsymbol{j}^-$/$\boldsymbol{v}^-$) are the forward (resp. backward) unidirectional fluxes.
For elementary reactions, the trajectories in the concentration species follows the chemical master equation. 
Taking the limit of large volume with fixed concentrations in the chemical master equation results in the kinetic rate equations written above in the deterministic regime \cite{McQuarrie1967, Gillespie1992, Kurtz1972}. 
Furthermore in this limit, the stochastic rates peak at their most probable value, yielding mass-action law for the unidirectional fluxes \cite{Ge2016, Pekar2005}:
\begin{align}
	\label{eq:MAL_def}
	j^{\pm \rho} = k^{\pm \rho} \, \prod_{s \hspace{.1em} \in \hspace{.1em} \E \cup \Z} [s]^{(\nabla_\pm)^s_\rho},
	&  & 
	v^{\pm r} = \nu^{\pm r} \, \prod_{e \in \E} [e]^{(\nabla_\pm^\E)^s_r}, 
\end{align}
where, $\rho \in \R$ and $r \in \nR$, and with $k^{\pm \rho}$/$\nu^{\pm r}$ the kinetic rate constants.

For simplicity, we will further assume that the mixture behaves ideally such that the chemical potential of species $s \in \Z \cup \E$ is
\begin{equation}
	\label{eq:mu_ideal}
	\mu_s = \mu^\circ_s + RT \ln \, [s].
\end{equation}
The local detailed balance condition relates the standard chemical potentials to the kinetic rate constants, ensuring thermodynamic consistency: 
\begin{align}
	\label{eq:LDB}
	\ln \dfrac{
		k^{+\rho}
	}{
		k^{-\rho}
	} 
	= - \dfrac{	
		\boldsymbol{\mu}^\circ \cdot \Nabla_\rho
	}{
		RT
	},
	& \quad &
	\ln \dfrac{
		\nu^{+r}
	}{
		\nu^{-r}
	} 
	= - \dfrac{	
		\boldsymbol{\mu}^\circ \cdot \Nabla_r 
	}{
		RT
	},
\end{align}
where $\Nabla_\rho$ (resp. $\Nabla_r$)  corresponds to the column associated with the autocatalytic reaction $\rho \in \R$ (resp. additional reaction $r \in \nR$) in the full stoichiometric matrix.

\NiceMatrixOptions{
	code-for-last-row = \color{black!70}\scriptstyle,
	code-for-first-col = \color{black!70}\scriptstyle,
}

\section{\label{Geometry}
	Non-equilibrium autocatalytic networks
}

\subsection{Conservation laws}

\subsubsection{Definition}

Conservation laws of a CRN are the row vectors $\boldsymbol{\ell}$ that belong to the left nullspace of the full stoichiometric matrix \cite{Rao2018, Rao2016, Avanzini2021}, 
\begin{equation}
	\boldsymbol{\ell} \cdot \Nabla = \boldsymbol{0}.
\end{equation}
Indeed, when the CRN is closed, $L = \boldsymbol{\ell} \cdot [\boldsymbol{s}]$ is a conserved quantity, \emph{i.e.} $\d_t L = 0$. 
Among the conservation laws, those that have only positive entries, $\ell_s \geq 0$ for all species $s$, are called \emph{mass-like} conservation laws. 
Lavoisier's law of mass conservation implies that the subset of mass-like conservation laws is never empty. 
In addition, there always exists a conservation law such that $\ell_s > 0$ for all species $s$, expressing the conservation of the total mass. 
While conservation laws are generally difficult to interpret, moieties form a subset of mass-like conservation laws that can be understood as the molecular fragments exchanged between species during chemical reactions \cite{Haraldsdottir2016}. 
For instance, in biochemical networks, common moieties include carbon conservation and phosphate group conservation (see also Fig.~\ref{fig:Glucose_cons_laws}\textbf{\textsf{a}}).

\subsubsection{Moiety matrix}

As a result, our stricter condition for autocatalysis Eq.~\eqref{eq:simpler_condtion} amounts to the absence of conservation law in the sub-network. 
The absence of conservation laws in $\S$ implies that they must be brought by the external species $\E$ to ensure mass conservation in the closed system ($\I = \boldsymbol{0}$). 
We can consider a linear basis of the conservation laws that we index by $i$, $\left\lbrace \boldsymbol{\ell}^i \right\rbrace$, .
We then define the matrix $\L = \left\lbrace\boldsymbol{\ell}^i \right\rbrace$, where the $i$-th row represents the conservation law $\boldsymbol{\ell}^i$.
As a result, $\L$ is a full-rank matrix satisfying $\L \cdot \Nabla = \boldsymbol{0}$.
Consequently, all the mass-like conservation laws of $\Nabla$ are either a row of $\L$ or can be obtained as a linear combination of the rows of $\L$.  

From the block decomposition Eq.~\eqref{eq:Nabla}, we have:
\begin{align}
	\L_\E \cdot \Nabla^\E_\R + \L_\Z \cdot \S = \boldsymbol{0}, \label{eq:L_R} & \quad &
	\L_\E \cdot \Nabla^\E_{\nR} = \boldsymbol{0}.
\end{align}
Now, as the rows of $\L$ are linearly independent, there exists a subset of external species $\Ep \subset \E$ such that the restriction of $\L$ to this subset, $\L_{\Ep}$, is (square) non-singular. 
Additionally, we denote $\Ef \subset \E$ as the remaining external species: $\E = \Ep \cup \Ef$.
Splitting Eq.~\eqref{eq:L_R} on these new subsets yields:
\begin{gather}
	\L_{\Ep} \cdot \Nabla^{\Ep}_{\R} +  \L_{\Ef} \cdot \Nabla^{\Ef}_{\R} + \L_{\Z } \cdot \S = \boldsymbol{0},  
\end{gather}	
and similarly for the additional reactions $\nR$. 
Then, from the non-singularity of $\L_{\Ep}$, we have:
\begin{gather} 
	\label{eq:Nabla_Ep_R} 
	\Nabla^{\Ep}_{\R} 
	= 
	- \M_{\Ef} \cdot \Nabla^{\Ef}_{\R} - \M_{\Z} \cdot \S,
\end{gather}
where we introduce the \emph{moiety matrix} of species $\Ep$ \cite{Avanzini2021}:
\begin{equation}
	\label{eq:M_def}
	\M = \left(\L_{\Ep}\right)^{-1} \cdot \L = \left( \mathbbm{1}_{|\Ep|}  ~,~ \M_{\Ef} ~,~ \M_{\Z} \right), 
\end{equation}
in which $\mathbbm{1}_{|\Ep|}$ stands for the $|\Ep| \times |\Ep|$ identity matrix.  
The moiety matrix is the row-reduced form of $\L$, which defines another linear basis for the conservation laws: $\M \cdot \Nabla = \boldsymbol{0}$.
Thus, with Eq.~\eqref{eq:M_def}, it becomes explicit that each conservation law (\emph{i.e.} each row of $\L$) is associated with a single species in $\Ep$ such that setting a non-vanishing external flux on species $e_p \in \Ep$ ($I^{e_p} \neq 0$) breaks its associated conservation law in $\L$. 
When the $i$-th row of $\L$, $\boldsymbol{\ell}^i$, corresponds to a moiety, its associated species in $\Ep$ represents the exchanged fragment, and the non-vanishing entries in the $i$-th row of $\M$ indicate where this fragment appears in the other species.
However, for conservation laws that cannot be interpreted as an exchanged fragment, the non-vanishing entries in $\M$ generalize the notion of a moiety.

\subsubsection{\label{geometry:broken_cons_laws}
	Broken conservation laws
}

Once all external species are coupled to the environment, their physical roles depend on whether they belong to $\Ep$ or to $\Ef$. 
Each species in $\Ep$ breaks its associated conservation law, while species in $\Ef$ induce flows through the system, effectively connecting external species.
As a result, species $\Ep$ remove constraints from the system, while species $\Ef$ enforce non-vanishing fluxes through the reactions.
From this, we can anticipate that if only potential species are present, the system cannot sustain non-vanishing fluxes, which prevents the network from settling into a non-equilibrium steady-state (NESS). 

Consequently, species $\Ep$ are referred to as the (external) \emph{potential} species; imposing a non-vanishing flux on all species in $\Ep$ breaks the conservation laws in $\L$.
Conversely, species $\Ef$ are referred to as external \emph{force} species: once coupled to the environment, they are capable of breaking detailed balance, thereby maintaining non-vanishing fluxes through the reactions \cite{Avanzini2021, Avanzini2022}.
Since the sub-network lacks conservation laws, coupling all external species to the environment results in breaking of all conservation laws of the full network. 
As a consequence, the matrix $\L$ directly provides the broken conservation laws of the open network, which play a pivotal role in the thermodynamics of the open system.

\vspace{.5\baselineskip}
\paragraph*{\textbf{Remark}} 
When the autocatalytic sub-network follows the most general criterion Eq.~\eqref{eq:stoichiometric_condition} instead of Eq.~\eqref{eq:simpler_condtion}, it is deprived of mass-like conservation laws. 
More specifically, the autocatalytic sub-network still possesses conservation laws that are non mass-like, \emph{i.e.} all of its conservation laws have entries of opposite sign. 
In that case, there might also be broken conservation laws in the sub-network originating from the autocatalytic species that are externally controlled.
When this arises, one should split the subset of externally controlled autocatalytic species $\Y$ into potential species $\Yp$, breaking the conservation laws of the sub-network, and force species $\Yf$. 

\begin{figure*}[!t]
	\centering
	\includegraphics[scale=.7]{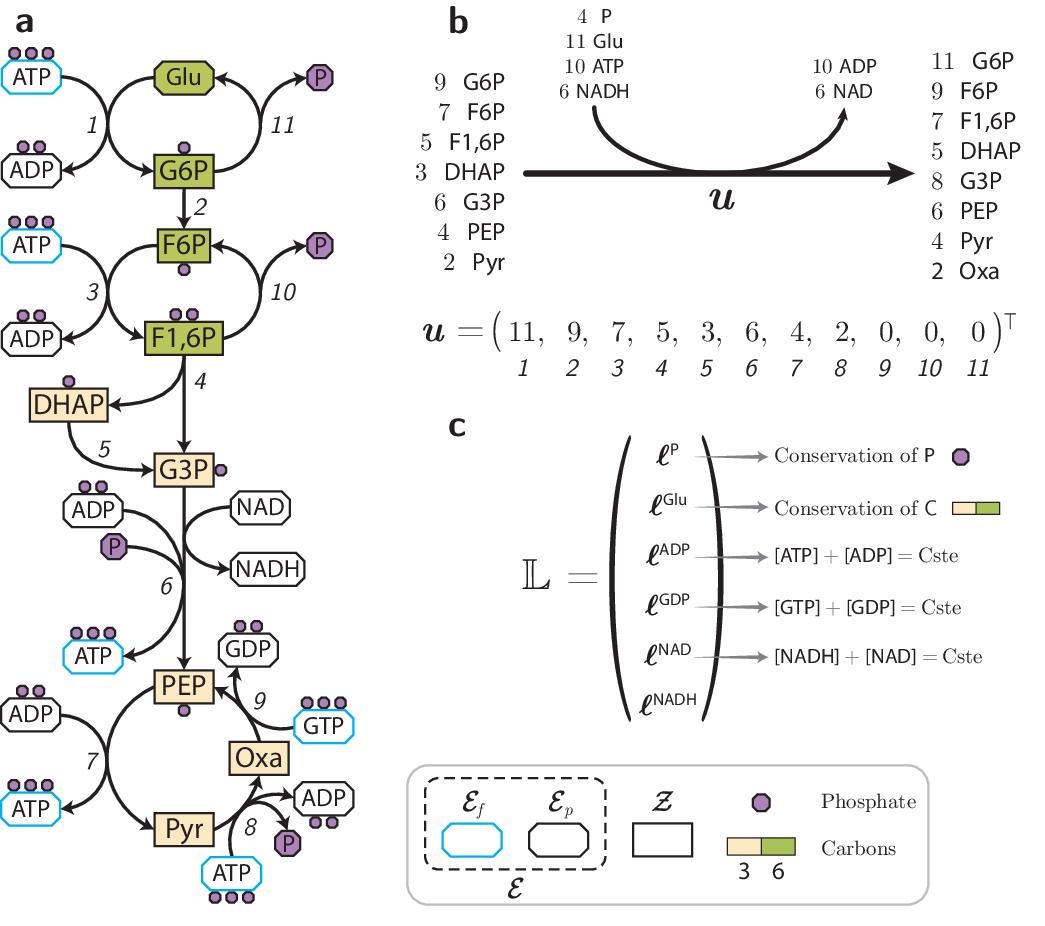}
	\caption{\label{fig:Glucose_cons_laws}
		Stoichiometric autocatalysis in glucose metabolism.  
		\textbf{\textsf{(a)}} 
		Simplified representation of glucose degradation into pyruvate (glycolysis) and its regulatory pathway (gluconeogenesis) adapted from Ref.~\cite{Lehninger2024} (Fig. 14-16). 
		In this network, the autocatalytic species $\Z$ are represented in the right angle boxes while the external species $\E$ are in the broken angle boxes. 
		For simplicity, we did not balance the reactions in $\textsf{H}^+$ and $\textsf{H}_2\textsf{O}$ and we omitted the enzymes catalyzing the reactions.
		\textbf{\textsf{(b)}}
		There exists a linear combination of the reactions, $\boldsymbol{u}$, that produces an excess of all the autocatalytic species $\Z$ at the expense of the external species, insuring the existence of an autocatalytic sub-network in the glucose metabolism. 
		\textbf{\textsf{(c)}}
		The external potential species $\Ep$ (black broken angle boxes) can be paired with the conservation law that they break once externally controlled. 
		In particular, carbon and phosphate group exchange among the species are two important moieties in the metabolism and are here represented using the color level and the purple octagons, respectively. 
		The remaining external species ($\textsf{ATP}$ \& $\textsf{GTP}$) are the external force species $\Ef$ that bring non-conservative forces that break detailed balance. 
	}
\end{figure*}

\subsubsection{Elementary modes of production}

From the stricter condition for autocatalysis in Eq.~\eqref{eq:simpler_condtion}, the rows of $\S$ are linearly independent, thus $\S$ admits a (non-unique) right-inverse $\invS$:
\begin{equation}
	\S \cdot \invS = \mathbbm{1}_{| \Z |}.
\end{equation}
The column associated with species $z \in \Z$ in $\invS$ is denoted $\boldsymbol{g}_z$. It defines a pathway that produces an excess of one unit of species $z$, leaving the other autocatalytic species unaffected. 
In other words, $\boldsymbol{g}_z$ defines an \emph{elementary mode of production} of species $z$, and, subsequently, the columns of $\invS$ are the productive modes of the autocatalytic sub-network.
Considering the sum of all the elementary modes, $\boldsymbol{u} = \sum_{z \in\Z} \boldsymbol{g}_z$, defines a pathway that produces an excess of one unit of \emph{all} the autocatalytic species: 
\begin{equation}
	\S \cdot \boldsymbol{u} = \boldsymbol{1} > 0.
\end{equation} 
As result, our stricter condition for autocatalysis is indeed sufficient to ensure that the stoichiometric criterion in Eq.~\eqref{eq:stoichiometric_condition} holds. 
Finally, as $\S$ is integer-valued, the productive modes can be always be rescaled, $\invS \to n  \invS$ for $n \in \mathbb{N}^\ast$, so that all the elementary modes are also integer-valued. 
Doing so, the (rescaled) elementary modes produce an excess of $n$ units of each autocatalytic species. 

\vspace{.5\baselineskip}
\paragraph*{\textbf{Remark}}
When the autocatalytic sub-network follows the most general criterion of Eq.~\eqref{eq:stoichiometric_condition}, the presence of (non mass-like) conservation laws prevents the definition of an elementary mode for each of the autocatalytic species. 
Instead, only a subset $\Z_f$ of autocatalytic species will be paired with elementary modes, and the production of the other autocatalytic species ($\Z_p = \Z - \Z_f$) will depend on the production of species in $\Z_f$.

\subsection*{Application to glucose metabolism}

As an example, we consider the network associated with glucose metabolism represented in Fig.~\ref{fig:Glucose_cons_laws}\textbf{\textsf{a}}. 
In the latter, glycolysis converts glucose into pyruvate with reactions 1 to 7. 
On the other hand, when pyruvate accumulates, gluconeogenesis allows to reform glucose from pyruvate with reactions 8 to 11. 
In this network, the external species are
\begin{equation*}
	\E = \left\lbrace \textsf{ATP,  ADP, GTP,  GDP,  NADH, NAD, P, Glu} \right\rbrace .
\end{equation*}
The external species are represented with broken angle boxes in Fig.~\ref{fig:Glucose_cons_laws}\textbf{\textsf{a}}. 
The remaining species of the network constitute the autocatalytic species set
\begin{equation*}
	\Z = \left\lbrace \textsf{G6P,  F6P,  F1,6P, DHAP,  G3P, PEP,  Pyr, Oxa} \right\rbrace,
\end{equation*}
and the autocatalytic species are represented with right angle boxes in Fig.~\ref{fig:Glucose_cons_laws}\textbf{\textsf{a}}.
All the reactions in Fig.~\ref{fig:Glucose_cons_laws}\textbf{\textsf{a}} belong to the set $\R$. 
As a result, there exists a linear combination of autocatalytic reactions $\boldsymbol{u}$ that produces \emph{all} the autocatalytic species from the external species, as represented in Fig.~\ref{fig:Glucose_cons_laws}\textsf{\textbf{b}}. 

In this network, there are 6 independent conservations laws which are all mass-like.
While, on the other hand, the autocatalytic sub-network follows our stricter condition and is deprived of conservation laws. 
A linear basis of the conservation laws can be constructed  such that one, $\boldsymbol{\ell}^{\, \textsf{P}}$, represents to the phosphate group exchanged between the species (represented with purples octagons in Fig.~\ref{fig:Glucose_cons_laws}\textbf{\textsf{a}}).  
Once the external species \textsf{P} is controlled, $\boldsymbol{\ell}^{\, \textsf{P}}$ is broken.
Another conservation law, $\boldsymbol{\ell}{}^{\, \textsf{Glu}}$, represents the exchange of carbon groups, which which is illustrated by the color background in the boxes of the species in Fig.~\ref{fig:Glucose_cons_laws}\textbf{\textsf{a}}. 
It is broken once glucose is controlled. 
Additionally, the total conservation of $\textsf{ATD/ADP}$,  $\textsf{GTP/GDP}$ and $\textsf{NADH/NAD}$ are represented by $\boldsymbol{\ell}{}^{\, \textsf{ADP}}$, $\boldsymbol{\ell}^{\, \textsf{GDP}}$ and $\boldsymbol{\ell}^{\, \textsf{NAD}}$. 
These are broken once $\textsf{APD}$, $\textsf{GDP}$ and $\textsf{NAD}$ are externally controlled. 
The last conservation law of the network is
\begin{equation}
	\boldsymbol{\ell}^{\, \textsf{NADH}} = ~~
	\begin{pNiceArray}{ccccccc}[last-row]
		1,  &   2,  &  2,  &  2,  &  2,  &  1,  & 1   \\ 
		\textsf{NADH}  & \textsf{Glu}  &  \textsf{G6P}  &  \textsf{F6P}  &  \textsf{F1,6P}  &  \textsf{G3P} & \textsf{DHAP}  
	\end{pNiceArray}
\end{equation}
(the entries associated with the missing species are zeros), which has no simple interpretation in terms of moiety, even though it is mass-like. 
It is broken once $\textsf{NADH}$ is exchanged with the environment.
The conservation laws define the rows of the matrix $\L$ (see also Fig.~\ref{fig:Glucose_cons_laws}\textbf{\textsf{c}}), and the potential species, which break the conservation laws, are 
\begin{equation*}
	\Ep = \left\lbrace \textsf{Glu, \ P,  \ ADP,  \ GDP, \ NAD, \ NADH} \right\rbrace. 
\end{equation*}
The remaining external species define the external force species, $\Ef = \left\lbrace \textsf{ATP, \ GTP}\right\rbrace$, and are represented with blue boxes in Fig.~\ref{fig:Glucose_cons_laws}\textbf{\textsf{a}}.

\subsection{\label{sec:cycles}
	Cycles
}

\subsubsection{Cycles of the network and the sub-network}

\begin{figure}[!ht]
	\centering
	\includegraphics{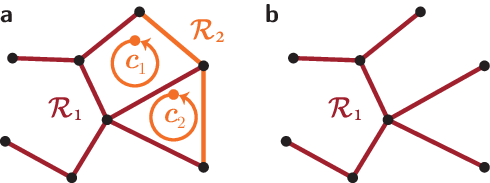}
	\caption{\label{fig:cyc_R1_R2}	
		Graph-theoretic interpretation of the subsets $\R_1$ and $\R_2$. 
		\textbf{\textsf{(a)}} 
		Each cycle $\boldsymbol{c}$ of a graph is associated with one reaction in the subset $\R_2$ (represented in orange) such that, by removing this reaction, the cycle $\boldsymbol{c}$ is removed as well. 
		\textbf{\textsf{(b)}} 
		Removing all the reactions $\R_2$ results in a graph deprived of cycles where the number of vertices corresponds to the number of edges. 
	}
\end{figure}

Cycles of the full CRN are the column vectors in the right nullspace of the full stoichiometric matrix: ${\Nabla \cdot \boldsymbol{\gamma} = \boldsymbol{0}}$. 
According to the block decomposition Eq.~\eqref{eq:Nabla} we have: 
\begin{gather}
	\NablaER \cdot \boldsymbol{\gamma}^\R + \NablaEnR \cdot \boldsymbol{\gamma}^{\nR} = \boldsymbol{0}, \label{eq:c_E} \\[0.5em]
	\S \cdot \boldsymbol{\gamma}^\R = \boldsymbol{0}, \label{eq:c_Z}
\end{gather}
where $\boldsymbol{\gamma}^R$ stands for the restriction of the cycle on the subset of reactions $R = \R$ or $\nR$.
Further, from Eq.~\eqref{eq:c_Z}, we see that if $\boldsymbol{\gamma}$ is a cycle, then either $\boldsymbol{\gamma}^\R = \boldsymbol{0}$, or $\boldsymbol{\gamma}^\R$ is itself a cycle of the autocatalytic sub-network and it belongs to the right-nullspace of $\S$. 
From the rank nullity theorem, there are $| \ker ~\S \ | = | \R | - | \Z |$ independent cycles in the sub-network. 
In what follows, we let $\left\lbrace \boldsymbol{c}_\varepsilon \right\rbrace_{1 \leq \varepsilon \leq |\R| - |\Z|} $ be a linear basis of these cycles.
As we did for the conservation laws, we define the matrix 
\begin{equation*}
	\mathbb{C} = \left\lbrace \boldsymbol{c}_\varepsilon \right\rbrace_{1 \leq \varepsilon \leq |\R| - |\Z|}, 
\end{equation*}
whose $\varepsilon$-th column is $\boldsymbol{c}_\varepsilon$. 
Subsequently, the matrix $\mathbb{C}$ is full-rank and verifies $\S \cdot \mathbb{C} = \boldsymbol{0}$. 

There exists a subset $\R_1 \subset \R$ of autocatalytic reactions such that the restriction $\S_{\R_1}$ is (square) non-singular, while $\R_2 = \R - \R_1$ stores the remaining autocatalytic reactions. 
By construction, $| \R_1 | = |\Z| $ while $|\R_2| = |\R| - |\Z| = | \ker ~\S \ |$ hence, each reaction in the subset $\R_2$ can be paired with one cycle of the sub-network such that, removing one reaction in $\R_2$ from the network causes precisely one cycle of $\S$ to disappear. 
Physically, removing one reaction that belongs to $\R_2$ removes exactly one cycle from the sub-network (see also Fig.~\ref{fig:cyc_R1_R2}\textbf{\textsf{a}}). 
Subsequently, if ones removes all the reactions in $\R_2$ from the hypergraph of the network, one ends up with the network $\S_{\R_1}$ which is then deprived of any cycle (see also Fig.~\ref{fig:cyc_R1_R2}\textbf{\textsf{b}}). 
Finally, the sub-matrix $\mathbb{C}^{\R_1}$ is directly related to $\mathbb{C}^{\R_2}$:
\begin{equation}
	\label{eq:c_R1}
	\mathbb{C}^{\R_1} = - \left( \S_{\R_1} \right)^{-1} \cdot \S_{\R_2} \cdot \mathbb{C}^{\R_2}.
\end{equation}
As the columns of $\mathbb{C}$ are linearly independent, this relation implies that $\mathbb{C}^{\R_2}$ is a (square) non-singular matrix.

\subsubsection{Emergent cycles}

We now examine the condition upon which a cycle of $\S$ is also a cycle when one considers the external species.
Indeed, a cycle of the sub-network will consume and/or produce external species. 
However, when a cycle of the sub-network preserves also the external species one has: 
\begin{equation}
	\label{eq:condition_cycle_0}
	\Nabla^\E_\R \cdot \boldsymbol{c} 
	= 
	\Nabla^\E_{\R_1} \cdot \boldsymbol{c}^{\R_1} + \Nabla^\E_{\R_2} \cdot \boldsymbol{c}^{\R_2} 
	= \boldsymbol{0}.
\end{equation}
Furthermore, we can use Eq.~\eqref{eq:c_R1} to replace $\boldsymbol{c}^{\R_2}$ such that Eq.~\eqref{eq:condition_cycle_0} becomes:
\begin{equation}
	\label{eq:condition_cycle}
	\left( \Nabla^\E_{\R_2} - \Nabla^\E_{\R_1} \cdot \left( \S_{\R_1} \right)^{-1} \cdot \S_{\R_2} \right) \cdot \boldsymbol{c}^{\R_2} = \boldsymbol{0}.
\end{equation}
In the left-hand side we recognize the Schur complement that we will denote
\begin{equation}
	\label{eq:effective_stoichiometric}
	\widetilde{\Nabla}{}^\E_{\R_2} \equiv \Nabla^\E_{\R_2} - \Nabla^\E_{\R_1} \cdot \left( \S_{\R_1} \right)^{-1} \cdot \S_{\R_2}. 
\end{equation} 
As a result, the right nullspace of the Schur complement, $ \mathrm{ker} ~ \widetilde{\Nabla}{}^\E_{\R_2} $, contains all of the cycles of the sub-network that also preserve the external species, and there is $| \mathrm{ker} ~ \widetilde{\Nabla}{}^\E_{\R_2} |$ independent such cycles.
When \emph{all} the cycles of $\S$ preserve also the external species, 
\begin{equation}
	\ker ~ \S = \ker ~ \Nabla_\R,
\end{equation}
one has $\widetilde{\Nabla}{}^\E_{\R_2} \cdot \mathbb{C}^{\R_2} = \boldsymbol{0}$. 
From the non-singularity of $\mathbb{C}^{\R_2}$, this occurs if, and only if, the Schur complement vanishes, $\widetilde{\Nabla}{}^\E_{\R_2} = \boldsymbol{0}$. 
 
In contrast, the cycles of the sub-network that are not also cycles of the full-network consume external species without production of autocatalytic species. 
In the CRN literature, they are labeled \emph{emergent cycles} as they exist only once all the external species are externally controlled. 
In metabolic networks, emergent cycles are also referred to as \emph{futile cycles} because they are useless for producing metabolites but are energy consuming (see next section).

\subsection{Dynamics}

\subsubsection{Linear basis of the elementary fluxes}

The right inverse $\mathbb{G}$ and the matrix $\mathbb{C}$ describe complementary effects for the sub-network: the columns of $\mathbb{G}$ are productive modes of the autocatalytic sub-network. 
In contrast, the columns of $\mathbb{C}$ are pathways preserving the state of the sub-network. 
To reflect this, we seek for productive modes that never proceed along the reactions in $\R_2$ by imposing $\mathbb{G}^{\R_2} = \boldsymbol{0}$. 
This constraint removes the non-uniqueness of $\mathbb{G}$ as the only right-inverse fulfilling this property is
\begin{equation}
	\label{eq:G_matrix}
	\mathbb{G} = \quad 
	\begin{pNiceArray}{m{1.5em}m{1.5em}}[last-row, first-col]
		\Vdots[line-style={solid, <->}, shorten=2pt]_{\small \rotatebox{90}{$\R_1$} }  & \Block{2-2}{(\S_{\R_1})^{-1}} & \\
		& & \\
		\hdottedline
		\Vdotsfor{2}[line-style={solid, <->}, shorten=2pt]_{\small \rotatebox{90}{$\R_2$} } \hspace*{1em} & \Block{2-2}{\boldsymbol{0}}  & \\
		&  & \\
		& \Hdotsfor{2}[line-style={solid, <->}, shorten=1pt]_{ \Z}  \\
	\end{pNiceArray}.
	\vspace*{0.5em}
\end{equation}
On top of that, we notice that the matrix $\left( \mathbb{G}, \, \mathbb{C} \right)$ has a non-vanishing determinant:
\begin{equation}
	\begin{vNiceArray}{m{1.5em}m{1.5em}:m{1.5em}m{1.5em}}[last-row]
		\Block{2-2}{(\S_{\R_1})^{-1}} &  & \Block{2-2}{\mathbb{C}^{\R_1}} &  \\
		&  &  & \\
		\hdottedline	
		\Block{2-2}{\boldsymbol{0}}  &  & \Block{2-2}{\mathbb{C}^{\R_2}} &  \\
		&  &  & \\
		\Hdotsfor{2}[line-style={solid, <->}, shorten=1pt]_{ |\R_1| } &  	\Hdotsfor{2}[line-style={solid, <->}, shorten=1pt]_{ |\R_2| }  \\
	\end{vNiceArray}
	= \dfrac{
		\mathrm{det} \left[ \mathbb{C}^{\R_2} \right] 
	}{
		\mathrm{det} \left[ \S_{\R_1} \right] 
	}
	\neq 0
	.
\end{equation}
Hence, the productive modes drawn from $\invS$ in Eq.~\eqref{eq:G_matrix} and the cyclic pathways in $\mathbb{C}$ define a linear basis of the flux space.

\subsubsection{Decomposition of the fluxes}

The chemical fluxes of the autocatalytic reactions, $\j$, can be decomposed on the linear basis defined by the columns of $\left( \mathbb{G}, \, \mathbb{C} \right)$:
\begin{align}
	\label{eq:decomposition_fluxes}
	\boldsymbol{j} 
	& = \mathbb{C} \cdot \boldsymbol{J} +  \mathbb{G} \cdot \boldsymbol{\J} \\[.5em]
	\label{eq:decomposition_fluxes_2} 
	& = \sum_\varepsilon J^{\varepsilon}(t) \, \boldsymbol{c}_\varepsilon
	+
	\sum_{z \in \Z} \J^{z}(t) \,  \boldsymbol{g}_{z},
\end{align}
where $J^{\varepsilon} (t)$ stands for the macroscopic flux along the cycle $\boldsymbol{c}_\varepsilon$ and $\J^{z} (t)$ is the macroscopic flux along the elementary mode $\boldsymbol{g}_{z}$. 
Furthermore, the constraint on the right inverse implies that the restriction of the fluxes on the subset of reactions $\R_2$ stems uniquely from the cycles of the sub-network:
\begin{equation}
	\label{eq:decomposition_j_R2}
	\boldsymbol{j}^{\R_2} 
	= \mathbb{C}^{\R_2} \cdot \boldsymbol{J} 
	= \sum_\varepsilon J^{\varepsilon}(t) \, \boldsymbol{c}_\varepsilon^{\R_2}.
\end{equation}
Hence, the macroscopic fluxes along the cycles of $\S$ are completely determined by the elementary fluxes of reactions in $\R_2$: 
\begin{equation}
	\boldsymbol{J} =  \left( \mathbb{C}^{\R_2} \right)^{-1} \cdot  \boldsymbol{j}^{\R_2}. 
\end{equation}
In contrast, the macroscopic fluxes in $\boldsymbol{\J}$ represent the rate at which the autocatalytic species are produced or consumed in the sub-network. 
Precisely, if the macroscopic flux associated with $z \in \Z$ is positive (resp. negative), $\mathcal{J}^z > 0$ (resp. $\mathcal{J}^z < 0$), the concentration of species $z$ increases (resp. decreases) in the sub-network as its elementary mode, $\boldsymbol{g}_{z}$, is performed forward (resp. backward).

Unlike the macroscopic fluxes along the productive modes, $\boldsymbol{\J}$, cannot be solely expressed as a function of $\j^{\R_1}$. 
Indeed, even though the fluxes along the cycles of $\S$ can be expressed solely with $\j^{\R_2}$ \emph{via} Eq.~\eqref{eq:decomposition_j_R2}, part of $\j^{\R_1}$ is also due to the cycles. 
Specifically, from Eq.~\eqref{eq:c_R1}, 
\begin{equation}
	- \left( \S_{\R_1} \right)^{-1} \cdot \S_{\R_2} \cdot \mathbb{C}^{\R_2} \cdot \boldsymbol{J} 
	=
	- \left( \S_{\R_1} \right)^{-1} \cdot \S_{\R_2} \cdot \j^{\R_2}
\end{equation}
is the contribution of the cycles in $\j^{\R_1}$.
As a consequence, $\boldsymbol{\J}$ is obtained by removing this contribution:
\begin{equation}
	\boldsymbol{\mathcal{J}} 
	=  
	\mathbb{S}_{\R_1} \cdot \boldsymbol{j}^{\R_1} 
	+ \S_{\R_2} \cdot \mathbb{C}^{\R_2} \cdot \boldsymbol{J}
	= \mathbb{S} \cdot \boldsymbol{j}.
\end{equation}

\subsubsection{Influx of external species}

Injecting the decomposition of the fluxes into the kinetic rate equations of the autocatalytic species provides: 
\begin{gather}
	\label{eq:production_flux_dt}
	\d_t [\z] = \boldsymbol{\J} + \I^{\Z}. 
\end{gather}
Similarly, by injecting the decomposition of the fluxes
into the rates equations of the external species we obtain:
\begin{equation}
	\label{eq:uptake_external_species}
	\d_t [\e] = \NablaER \cdot \invS \cdot \boldsymbol{\J} ~+~ \NablaER \cdot \mathbb{C} \cdot \boldsymbol{J} ~+~ \NablaEnR \cdot \boldsymbol{v} ~+~ \I^\E.
\end{equation}
The first term in RHS corresponds to the influx of external species needed to produce (or consume) autocatalytic species using the productive modes, using Eq.~\eqref{eq:production_flux_dt} it can be expressed as 
\begin{equation}
	\label{eq:influx_production}
	\NablaER \cdot \invS \cdot \boldsymbol{\J} = \d_t \Delta \e (\z) - \NablaER \cdot \invS \cdot \I^{\Z}.
\end{equation}
Hence the uptake of external species needed for producing an excess $\Delta \z = [\z] (t) - [\z] (0)$ of autocatalytic species is
\begin{equation}
	\label{eq:Delta_e_def}
	\Delta \e (\z) = \NablaER \cdot \invS \cdot \Delta \z,   
\end{equation}
which depends only on the excess produced.
In contrast, the second term in the RHS of Eq.~\eqref{eq:uptake_external_species} quantifies the uptake of external species along the cycles of the sub-network. 
Using Eq.~\eqref{eq:c_R1}, it can be written  with the Schur complement:
\begin{equation}
	\label{eq:influx_ex_cyc}
	\NablaER \cdot \mathbb{C} \cdot \boldsymbol{J}
	=
	\widetilde{\Nabla}{}^\E_{\R_2} \cdot \mathbb{C}^{\R_2} \cdot \boldsymbol{J}.
\end{equation}
Because the cycles of the sub-network that are also cycle of the full network belong to the right-nullspace of the Schur complement, these cycles do not contribute in Eq.~\eqref{eq:influx_ex_cyc} where only the emergent/futile cycles remain. 
In particular, if $\ker~ \S = \ker  \Nabla_\R$, all the cycles of the sub-network are also preserving the external species and, in that case, the Schur complement vanishes, and so it is for Eq.~\eqref{eq:influx_ex_cyc}. 
Finally, the last two terms in Eq.~\eqref{eq:uptake_external_species} are, respectively, the change of external species along the additional reactions ($\nR$) and the fluxes coupling the external species with the environment.

Hence, Eq.~\eqref{eq:influx_production} highlights that the uptake of external species along the productive modes derived from an exact time derivative. 
As a result, it will be independent of the specific pathway taken by the autocatalytic species in the sub-network. 
Hence, reminiscent of Newtonian mechanics, we will refer to the influx of external species fueling the production of autocatalytic species as the \emph{conservative} influx.
As we shall see in the next section, this analogy can be pushed further as the chemical forces driving this process can be derived from a potential. 
The latter will be essential to study the thermodynamics of an autocatalytic sub-network. 
On the contrary, the uptake of external species along the cycles of the sub-network cannot be written as an exact time derivative, and, as a result, they will bring non-conservative forces.

\subsubsection{Steady fluxes}

From Eq.~\eqref{eq:dynamics_subnetwork}, steady-state in the open sub-network is achieved when the concentrations of the autocatalytic species that are not externally coupled, \emph{i.e.} the $\X$ species, become constant in time. 
This situation occurs when the fluxes verify $\overline{\j} \in \mathrm{ker} ~\S{}^{\, \X}$ \cite{Qian2005, Polettini2014}. 
Clearly, all the cycles of $\S$ remain cycles for $\S{}^{\, \X}$ and, then, the columns of $\mathbb{C}$ defines a family of linearly independent vector which generates a subset of $\mathrm{ker} ~\S{}^{\, \X}$. 
Additionally, the elementary modes of the species $\Y$ let each of the $\X$ species unaffected by definition.
As a result, the columns of $\invS_\Y$ also generate subset of $\mathrm{ker} ~\S{}^{\, \X}$ and are linearly independent from the columns of $\mathbb{C}$.
Finally, the rank-nullity theorem implies that $\left( \mathbb{C}, \ \invS_\Y \right)$ is a linear basis of $\mathrm{ker} ~\S{}^{\, \X}$.
Then, at steady-state, the decomposition of the fluxes becomes:
\begin{align}
	\label{eq:ss_fluxes}
	\overline{\j} & = \mathbb{C} \cdot \overline{\boldsymbol{J}} + \invS_{\Y} \cdot \overline{\boldsymbol{\J}}{}^{\,\Y}. 
\end{align}

Removing or adding intermediates species by the means of one-to-one reactions (fueled with an arbitrary number external species as reactants and products) is an important class of coarse-graining that preserves the decomposition of the fluxes. 
This procedure is, for example, used to derive the Michaelis-Menten kinetics or the Hill kinetics from elementary steps \cite{Wachtel2018}. 
Aside from adding/removing linear pathways, more advanced reductions preserving steady-state have been recently studied \cite{Sughiyama2022}. 
Finally, this decomposition serves as the starting point to establish a circuit theory of open chemical networks, drawing an analogy between CRNs and electronic circuits \cite{Avanzini2023, Raux2024}. 

\begin{figure}[!t]
	\centering
	\includegraphics[scale=.8]{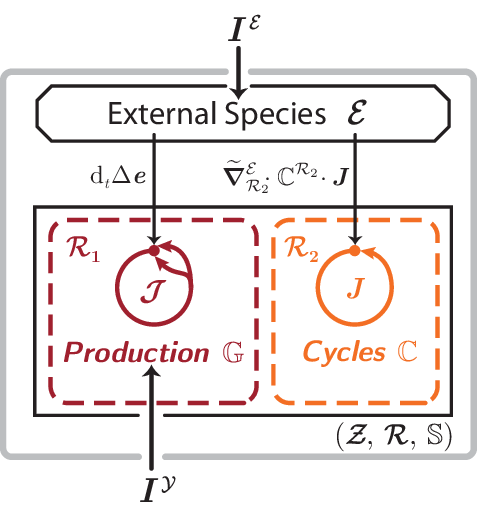}
	\caption{\label{fig:schema_system}
		Schematic representation of the system Eq.~\eqref{eq:Nabla}. 
		To fulfill mass-conservation, the autocatalytic sub-network, represented in the right angle box, is coupled with external species.
		The autocatalytic reactions, $\R$, splits into two disjoint subsets.
		The reactions in $\R_1$ convert $\Delta \e (\z)$ external species $\E$ into an excess of autocatalytic species $\Z$. 
		While the reactions in $\R_2$ are associated with the cycles of $\S$. 
		These are sustained thanks to a non-conservative influx of external species. 
		All the external species and the $\Y$ species can be exchanged with the environment, which is delimited by the gray box, by the mean of external fluxes. 
		Finally, the production is sustained \emph{via} the autocatalytic species exchanged with the environment.
	}
\end{figure}

\subsubsection{Physical role of the species}

As a direct consequence of Eq.~\eqref{eq:ss_fluxes}, when only the external species are externally controlled (and $\Y = \varnothing$), the steady-fluxes align with the cycles of the sub-network. 
Among these cycles, the emergent cycles are the ones consuming the external species and contributing to Eq.~\eqref{eq:influx_ex_cyc}. 
As a result, maintaining fluxes along the emergent cycles require to exert non-conservative forces on the sub-network.
Additionally, when only the external species are controlled, only the external force species $\Ef$ bring such forces in the network. 
Hence, we can relate the number of external force species $|\Ef|$ to the number of emergent cycles in the sub-network (see Appendix C for the proof),
\begin{equation}
	\label{eq:cardinal_Ef}
	\begin{aligned}
		| \Ef | & = 
		| \nR | 	
		+ | \R_2 |  	
		- |\mathrm{ker} ~ \widetilde{\Nabla}{}^\E_{\R_2} |, 
	\end{aligned} 
\end{equation}
from which the number of conservation laws brought by the external species follows: 
\begin{equation}
	|\Ep| = |\E| - |\Ef|. 
\end{equation}
In other words, Eq.~\eqref{eq:cardinal_Ef} states that the physical role of the external force species is (i) to maintain all the additional reactions away from equilibrium and (ii) to sustain non-vanishing fluxes along the 
$| \R_2 | - |\mathrm{ker} ~ \widetilde{\Nabla}{}^\E_{\R_2} |$
emergent cycles of the sub-network. 

Exerting non vanishing external fluxes on the $\Y$ species caused $| \Y |$ new cycles to exist for the remaining $\X$ species. 
These are all \emph{emergent} as they do not exist prior to controlling the $\Y$ species. 
On the other hand, as they do not break any conservation law, the $\Y$ species are "force" species that maintain the flux along their respective elementary mode.
Indeed, to sustain steady production (or consumption) of the autocatalytic species $y$, the latter should be coupled with the environment by the means of a non-vanishing external flux, $I^y \neq 0$, such that $\J^y (t) \xrightarrow{\hspace{1em}} \overline{\J}{}^{\, y} = - I^{\, y}$.
Finally, the macroscopic fluxes along the $\X$ species tend to vanish as NESS settles in the sub-network, 
$ \boldsymbol{\J}{}^\X (t) \longrightarrow \boldsymbol{0}$. 
Nevertheless, during the transient dynamics, they take non-zero values, and serve as the relaxation modes of the sub-network needed to reach steady-state \cite{Cengio2022}.

\subsection*{Application to glucose metabolism II}

\begin{figure*}[!t]
	\includegraphics[scale=.7]{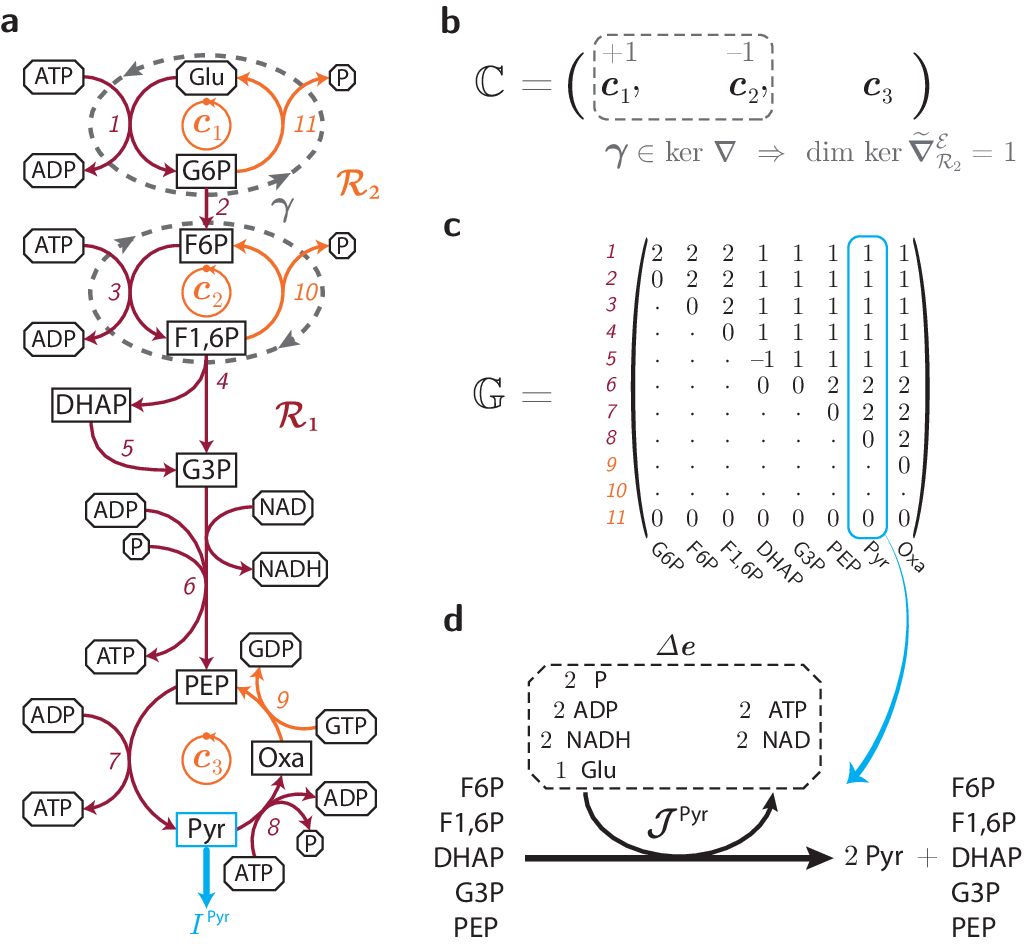}
	\caption{\label{fig:Glucose_Dynamic}
		Nonequilibrium dynamics of glucose metabolism. 
		\textbf{\textsf{(a)}} 
		The reactions associated with glycolysis and the conversion of pyruvate into oxaloacetate of gluconeogenesis (reaction $\mathsfit{8}$) define the subset $\R_1$, represented with solid red arrows. 
		The other reactions of gluconeogenesis bring cycles in the sub-network and define the subset $\R_2$, represented with solid orange arrows. 
		\textbf{\textsf{(b)}}
		A linear basis of the cycles of the sub-network is formed by these three cycles. 
		Conversely, the full network has only one cycle (represented with the gray dotted lines in \textbf{\textsf{a}}), formed by a linear combination of two cycles of the sub-network. 
		As a consequence, the cycles shared by the full-network and the sub-network is a linear space of dimension one.
		\textbf{\textsf{(c)}} 
		The subset $\R_1$ allows to define independent pathways producing each of the autocatalytic species that are linearly independent from the cycles. 
		\textbf{\textsf{(d)}}
		Setting a non-vanishing external flux on the pyruvate one selects its elementary mode at steady-state, recovering the conversion of one unit of glucose into two units of pyruvate that regenerates \textsf{ATP} from \textsf{ADP} \cite{Wachtel2022}. 
	}
\end{figure*}

In glucose metabolism, there are three more reactions than the number of autocatalytic species ($|\R| = 11$ and $|\Z| = 8$), as a result, rank-nullity theorem imposes that there are three independent cycles in the autocatalytic sub-network:
\begin{align}
	\boldsymbol{c}_1 = \begin{pNiceArray}{w{c}{.5em}w{c}{.5em}}[last-row]
		1,  &   1  \\ 
		\sf{1}  &  \sf{11}  
	\end{pNiceArray}^\top,
	& & 
	\boldsymbol{c}_2 = \begin{pNiceArray}{cc}[last-row]
		1,  &   1  \\ 
		\sf{3}  &  \sf{10}  
	\end{pNiceArray}^\top,
	&  & 
	\boldsymbol{c}_3 = \begin{pNiceArray}{ccc}[last-row]
		1,  &   1,  & 1\\ 
		\sf{7}  &  \sf{8}  &  \sf{9}  
	\end{pNiceArray}^\top,
\end{align}
where the entries in the remaining reactions vanish. 
These cycles are caused by the inhibitory reactions of gluconeogenesis, $\R_2 = \left\lbrace \sf{9, \, 10, \, 11}\right\rbrace$, which are represented with orange arrows in Fig.~\ref{fig:Glucose_Dynamic}\textbf{\textbf{a}}. 
Removing these reactions results in a network deprived of cycles, where only the reactions in $\R_1$ remain (red arrows in Fig.~\ref{fig:Glucose_Dynamic}\textsf{\textbf{a}}).
These three cycles define the columns of the matrix $\mathbb{C}$.
Taking the external species into account, the full network has one cycle $\boldsymbol{\gamma}$ (gray dashed arrows in Fig.~\ref{fig:Glucose_Dynamic}\textsf{\textbf{a}}) that is formed by a linear combination of two cycles of the sub-network: 
\begin{equation}
	\boldsymbol{\gamma} = \boldsymbol{c}_1 - \boldsymbol{c}_2.
\end{equation}
Hence, the kernel of the Schur complement is of dimension one (see also Fig.~\ref{fig:Glucose_Dynamic}\textsf{\textbf{b}}).
Thus, when glycolysis and gluconeogenesis are considered simultaneously, two independent emergent/futile cycles exist. 
These two degrade \textsf{ATP}/\textsf{GTP} into \textsf{ADP}/\textsf{GDP} without any useful production of metabolite: 
\begin{equation*}
	\begin{gathered}
		\ce{ \textsf{ATP} ->[\boldsymbol{c}_2] \textsf{ADP} + \textsf{P} }, \\[.5em]
		\ce{ 2 \ \textsf{ATP} + \textsf{GTP} ->[\boldsymbol{c}_3] 2 \ \textsf{ADP} + \textsf{GDP} + 3 \ \textsf{P} }.
	\end{gathered}
\end{equation*}
On the other hand, the reactions in $\R_1$ are sufficient to define elementary modes for the autocatalytic species that are linearly independent of the three cycles of the sub-network. 
These elementary modes are the columns of the matrix $\invS$: each column of $\invS$ defines an overall reaction that produces an excess of its associated autocatalytic species, see Fig.~\ref{fig:Glucose_Dynamic}\textbf{\textsf{c}}. 
Consequently, the fluxes of the autocatalytic reactions, $\j$, can be written on the basis defined by $\left( \mathbb{C}, \, \invS \right)$ according to Eq.~\eqref{eq:decomposition_fluxes}.
When pyruvate is externally controlled, the dynamics tend to align the fluxes along its mode, $\boldsymbol{g}_\textsf{Pyr}$, such that, at steady-state,
\begin{equation}
	\overline{\j} ~=~ \sum_{\varepsilon = 1}^3 \overline{J}{}^{\, \varepsilon} \, \boldsymbol{c}_\varepsilon ~+~ \overline{\J}{}^{\, \textsf{Pyr}} \, \boldsymbol{g}_\textsf{Pyr}.
\end{equation}
This results in the well-known conversion of one glucose unit into two units of pyruvate that replenishes $\textsf{ATP}$ from $\textsf{ADP}$, as represented in Fig.~\ref{fig:Glucose_Dynamic}\textbf{\textsf{d}}.

Here, one cycle is shared between the autocatalytic sub-network and the full network ($\boldsymbol{\gamma} = \boldsymbol{c}_1 - \boldsymbol{c}_2$), hence Eq.~\eqref{eq:cardinal_Ef} imposes that there should be two external force species in order to maintain the fluxes along the emergent cycles of the sub-network. 
Finally, we note that the mode of production of \textsf{DHAP} requires to perform reaction $\mathsfit{5}$ backward. 
In this case, it is admissible as this reaction is indeed reversible. 
If this was not the case, the macroscopic flux along this mode would always vanish.

%
%
\section{\label{Thermo}
	Thermodynamics of the autocatalytic sub-network
}

\subsection{Fundamental non-conservative forces}

The total entropy production rate (EPR) quantifies the amount of energy dissipated by the reactions in the system. 
It can be expressed as the contribution of the reactions of the sub-network ($\R$) and the additional reactions ($\nR$):
\begin{equation*}
	\dot{\Sigma} = \dot{\Sigma}_\R + \dot{\Sigma}_{\nR} \geq 0. 
\end{equation*}
Starting from the chemical master equation that governs the stochastic trajectories in the concentration space, and taking the large volume limit, ends up in the following form for the dissipation of the dissipation along the reactions of the sub-network in the deterministic regime \cite{Ge2016}:
\begin{equation}
	T \dot{\Sigma}_{\R} = \sum_{\rho \in \R} (j^{+\rho} - j^{- \rho}) \ \ln \left( \dfrac{j^{+ \rho}}{j^{- \rho}} \right),
 \end{equation}
expressed in units where $RT = 1$. 
In the dissipation, the log-ratio of the unidirectional fluxes define the chemical forces acting on the reactions, \emph{i.e.} the affinities
\begin{equation}
	\label{eq:flux_force}
	\mathcal{A}_\rho \equiv \ln \left( \dfrac{j^{+ \rho}}{j^{- \rho}} \right) = - \left(  \Mu \cdot \Nabla \right) _\rho =- \Delta G_\rho,
\end{equation}
which correspond to the opposite of the Gibbs free-energy differences of the reactions. 
Using Eq.~\eqref{eq:Nabla_Ep_R} allow us to replace the stoichiometric matrix of the external potential species, $\Nabla^{\Ep}_\R$, in Eq.~\eqref{eq:flux_force}. 
By doing so, we introduce the following gauge transform of the chemical potentials:
\begin{equation}
	\label{eq:chemical_force}
	\F = \Mu - \Mu_{\Ep} \cdot \mathbb{M}, 
\end{equation}
such that the affinities of the reactions can be written as 
\begin{equation}
	\label{eq:A_vs_F}
	\mathcal{A}_\rho = - \left( \F \cdot \Nabla \right)_\rho.
\end{equation}
As a result, when $\F = \boldsymbol{0}$, all the affinities vanish implying that the sub-network have reached detailed balance. 
Hence, the entries in $\F$ are the \emph{fundamental non-conservative forces} \cite{Avanzini2021, Rao2018} associated with each of the chemical species. 
As anticipated, by definition, these forces necessarily vanish for the potential species: $\F_{\Ep} = \boldsymbol{0}$. 
 
Finally, the dissipation along the autocatalytic reactions can be written as a function of the non-conservative forces of the species:
\begin{equation}
	T \dot{\Sigma}_{\R}  =  - \F \cdot \Nabla_{\R} \cdot \j. 
\end{equation}
Similarly, the dissipation along the additional reactions reads
\begin{equation}
	T \dot{\Sigma}_{\nR}  =  - \F \cdot \Nabla_{\nR} \cdot \boldsymbol{v} = - \F_\E \cdot \Nabla_{\nR}^\E \cdot \boldsymbol{v}. 
\end{equation}

\subsection{Unconditionally detailed balanced sub-networks}

From the decomposition of steady-fluxes Eq.~\eqref{eq:ss_fluxes}, when autocatalytic species are exchanged with the environment ($\Y \neq \varnothing$), 
macroscopic currents flow along the productive modes of species $\Y$ at steady-state, and reaching a detailed balance  (equilibrium) in the sub-network is impossible. 
In contrast, when $\Y = \varnothing$, the sub-network might still be able to reach equilibrium for arbitrary values of the chemical potentials of external species $\Mu_\E$. 
When this occurs, the sub-network is said to be unconditionally detailed balanced \cite{Avanzini2021} and the chemical potentials of the free autocatalytic species $\Mu_{\X} = \Mu_{\Z}$, are able to balance arbitrary $\Mu_\E$:
\begin{align}
	\Mu_{\Z}^\text{eq} \cdot \S
	=
	- \Mu_\E \cdot \NablaER. 
\end{align}
As $\S_{\R_1}$ is non-singular, the unique candidate for $\Mu_{\Z}^\text{eq}$ is
\begin{equation}
	\label{eq:equilibrium_state_mu}
	\Mu_{\Z}^\text{eq} = - \Mu_\E \cdot \Nabla^\E_{\R_1} \cdot \left( \S_{\R_1} \right)^{-1}.
\end{equation}
Plugging this solution in the equations for reactions $\R_2$ provides the feasibility of this candidate, namely:
\begin{equation}
	\label{SM-eq:consistency_condition}
	\Mu_\E \cdot \widetilde{\Nabla}{}^\E_{\R_2} 
	= \boldsymbol{0}.
\end{equation} 
In light of the previous chapter, unconditional detailed balance is achieved if, and only if, the Schur complement vanishes, \emph{i.e.} when
\begin{equation*}
	\ker~ \S  = \ker~ \Nabla_\R.
\end{equation*}
Importantly, this condition is independent of the presence of non-conservative forces brought by the external force species, $\F_{\Ef}$, and relies solely on topology. 
This is expected from Eq.~\eqref{eq:cardinal_Ef} because, when the Schur complement vanishes, species $\Ef$ serve only to maintain fluxes along the additional reactions $\nR$.

\subsection{Thermodynamical potential}

The Gibbs free-energy of the closed system ($\I = \boldsymbol{0}$) is well-known:
\begin{equation}
	\label{eq:close_G}
	G(\z, \ \e) = \sum_{e \in \E} [e] \left( \mu_e - 1 \right) + \sum_{z \in \Z} [z] \left( \mu_z - 1 \right),
\end{equation}
where the "$-1$" is the contribution of the solvent.  
By setting off the non-vanishing external fluxes on the external species and on species $\Y$, the system starts exchanging moieties with its environment:
\begin{equation}
	\boldsymbol{m}(\z, \e) = \M_\E \cdot [\e] + \M_\Z \cdot [\z].
\end{equation}
Hence, in the presence of non-vanishing external fluxes, 
\begin{equation}
	\label{eq:G_def}
	\G(\z, \e) = G(\z, \e) - \Mu_{p} \cdot \boldsymbol{m}(\z, \e) + \F_{\E} \cdot \Delta \e (\z), 
\end{equation}
corresponds to the thermodynamical potential of the open sub-network which captures its thermodynamics behavior. 
The first two terms in the definition of $\G$ correspond to the semigrand free energy of an open CRN whcih was derived previously for stochastic \cite{Rao2018} or deterministic open chemical networks \cite{Rao2016, Avanzini2021, Avanzini2022}.
Yet, the semigrand free-energy fails to predict relaxation towards equilibrium for unconditionally detailed balance sub-networks in the presence of additional reactions. 
To fix this, we need to take into account that the non-conservative forces of the external species do not exert along the productive modes of the sub-network because the influx of external species along the productive modes of the sub-network is conservative.  
This is implemented by the last therm in Eq.~\eqref{eq:G_def}, $\F_\E \cdot \Delta \e $. 
 
\vspace*{1em}

\subsection{Decomposition of the EPR}

The time evolution of $\G$ can be written as:
\begin{widetext}
\begin{equation}
	\label{eq:evol_G_1}
	\d_t \G = - T \dot{\Sigma} 
	- \d_t \Mu_{\Ep} \cdot \boldsymbol{m} 
	+ \d_t \F_{\E} \cdot  \Delta \e 
	+ \F_{\E} \cdot \left( \I^{\E} + \d_t \Delta \e \right) 
	+ \F_\Z \cdot \I^\Z. 
\end{equation}
\end{widetext}
Gathering the terms in the RHS of Eq.~\eqref{eq:evol_G_1} as follows:
\begin{gather}
	\label{eq:W_driv}
	\dot{W}_\text{driv} =  - \d_t \Mu_{\Ep} \cdot \boldsymbol{m} + \d_t \F_{\E} \cdot  \Delta \e, \\[0.5em]
	\label{eq:W_nc_E}
	\dot{W}_{\text{nc}, ~ \E} = \F_{\E} \cdot \left( \I^{\E} + \d_t \Delta \e \right), \\[0.5em]
	\label{eq:W_nc_Z}
	\dot{W}_{\text{nc}, ~ \Z} =  \F_{\Z} \cdot \I^\Z,
\end{gather}
allows to write the change in free-energy in a thermodynamically appealing way
\begin{equation}
	\label{eq:second_law_dt_G}
	\d_t \G = - T \dot{\Sigma} + \dot{W}_\text{driv} + \dot{W}_{\text{nc}, ~ \E} + \dot{W}_{\text{nc}, ~ \Z}.
\end{equation}
In this decomposition, $\dot{W}_\text{driv}$ represents the driving work rate required to change the equilibrium state in the autocatalytic sub-network by varying the chemical potential of the potential species.
Because the equilibrium state of the sub-network depends on all the external species, the external force species also contribute to the driving work.  
The two non-conservative work rates, $\dot{W}_{\text{nc}, ~ \E}$ and $\dot{W}_{\text{nc}, ~ \Z}$ describe the power performed by species $\E$ and $\Z$ tho maintain to maintain non-vanishing fluxes.
Hopefully, the potential species does not contribute to $\dot{W}_{\text{nc},\ \E}$.

\subsection{Autonomous autocatalytic networks}

When the driving rate vanishes,
\begin{equation}
	\dot{W}_\text{driv} = 0,
\end{equation}
the sub-network is said to be autonomous.
This is implemented by fixing the chemical potentials of the external species to a constant value in the system through external particle reservoirs (chemostats). 
Autonomy can also be achieved if there exists a timescale separation between the dynamics of the external species and the one of the autocatalytic sub-network such that, in the time scale of the sub-network, the chemical potentials of the external species can be regarded as constant. 
In that case, the external fluxes on the external species $\I^\E$ should balance the consumption/production of external species by the chemical reactions in Eq.~\eqref{eq:uptake_external_species}. 

\subsubsection{Thermodynamic cost of production}

For autonomous sub-networks, the decomposition of the EPR can be written as:
\begin{equation}
	\label{eq:second_law_autonomous}
	\d_t \G 
	= 
	- T \dot{\Sigma}_{\R} + \dot{W}_\text{cyc} + \wprod. 
\end{equation}
In this decomposition,
\begin{equation}
	\dot{W}_\text{cyc}
	= - \F_\E \cdot \widetilde{\Nabla}{}^\E_{\R_2} \cdot \mathbb{C}^{\R_2} \cdot \boldsymbol{J},
\end{equation}
accounts for the power needed to sustain the emergent/futile cycles of $\S$. 
Furthermore, the total work rate sustaining the cycles can be further decomposed on each cycle $\boldsymbol{c}_\varepsilon$ of the sub-network. 
Indeed, as from Eq.~\eqref{eq:A_vs_F} $- \F \cdot \Nabla = \boldsymbol{\mathcal{A}}$, the $\varepsilon$-th entry in 
\begin{equation*}
	- \F_\E \cdot \widetilde{\Nabla}{}^\E_{\R_2} \cdot \mathbb{C}^{\R_2} = - \F \cdot \Nabla_\R \cdot \mathbb{C} = \boldsymbol{\mathcal{A}} \cdot \mathbb{C}
\end{equation*}
is $\mathcal{A} [ \boldsymbol{c}_\varepsilon ] = - \Delta G [\boldsymbol{c}_\varepsilon ]$, the overall affinity associated with cycle $\boldsymbol{c}_\varepsilon$ which is the opposite of its free-energy difference. 
As a result, the power maintaining the cycles can be written as
\begin{equation}
	\label{eq:w_cyc}
	\dot{W}_\text{cyc} =  \sum_\varepsilon \mathcal{A} [ \boldsymbol{c}_\varepsilon ] \ J^\varepsilon = - \sum_\varepsilon \Delta G [\boldsymbol{c}_\varepsilon ] \ J^\varepsilon,
\end{equation}
such that the $\varepsilon$-th summand corresponds to the specific power allocated to cycle $\boldsymbol{c}_\varepsilon$, which vanishes if $\boldsymbol{c}_\varepsilon$ is also a cycle of the full network because its affinity vanishes in that case. 
Consequently, only the emergent/futile cycles will require energy to be maintained while producing no net excess of autocatalytic species. 

In addition,
\begin{equation}
	\wprod =  \left( 
	\F_\E \cdot \Nabla^\E_{\R_1} \cdot (\S_{\R_1})^{-1}  + \F_{\Z}
	\right) \cdot \I^{\Z}
\end{equation}
represents the power needed to sustain the production of the autocatalytic species $\Z$ along their elementary modes. 
Its first term, 
\begin{equation*}
	\F_\E \cdot \Nabla^\E_{\R_1} \cdot (\S_{\R_1})^{-1} \cdot \I^{\Z},
\end{equation*}
corresponds to the uptake of external species fueling the elementary modes of production.
As the fluxes decomposition suggested, only the species $\Y$ contribute to $\dot{W}_\text{prod}$:
\begin{equation}
	\label{eq:w_prod}
	\wprod = 
	- \sum_{y \in \Y}  \mathcal{A} [ \boldsymbol{g}_{y} ] \ I^{y} 
	=  
	\sum_{y \in \Y}  \Delta G  [ \boldsymbol{g}_{y} ] \ I^{y}
\end{equation}
where 
\begin{equation}
	\mathcal{A} [ \boldsymbol{g}_{y} ] = - \F \cdot \Nabla_\R \cdot \boldsymbol{g}_{y} = - \F_\E \cdot \Nabla_\R^\E \cdot \boldsymbol{g}_{y} - \mathcal{F}_{y}
\end{equation}
is the overall affinity along the elementary mode of species $y$, $\boldsymbol{g}_{y}$, which is the opposite of its Gibbs free-energy difference $ \Delta G  [ \boldsymbol{g}_{y} ] $. 
Similarly to the cycles, the $y$-th summand in Eq.~\eqref{eq:w_prod} is the specific work rate dedicated to the production of species $y$.

\subsubsection{Unconditionally detailed balanced sub-networks}

From the EPR decomposition in Eq.~\eqref{eq:second_law_autonomous}, it is clear that whenever $\dot{W}_\text{cyc} = 0$ and $\dot{W}_\text{prod} = 0$, the $\G$ tends converges to its minimum $\d_t \G = - T \dot{\Sigma}_\R \leq 0$.
These conditions are equivalent to
\begin{align*}
	\widetilde{\Nabla}{}^{\, \E}_{\R_2} = \boldsymbol{0}
	& & \text{and} & & 
	\Y = \varnothing.
\end{align*}
When this occurs, the sub-network is unconditionally detailed balance. 
Furthermore, when these two conditions are fulfilled, the optimum of $\G$ with respect to autocatalytic species $\Z = \X$ is attained when
\begin{equation}
	\label{eq:equilibrium_from_G}
	\Mu_{\Z}^\text{eq} = - \Mu_{\E} \cdot \Nabla^\E_{\R_1} \cdot \left( \S_{\R_1} \right)^{-1},
\end{equation}
which is in accordance with what we found by directly balancing the chemical potentials in Eq.~\eqref{eq:equilibrium_state_mu}. 
Additionally, the equilibrium state is the global minimum of $\G$: 
\begin{equation}
	\label{eq:DeltaG_1}
	\begin{aligned}
		\Delta \G & = \G(\z, \, \e) - \G(\z^\text{eq}, \, \e) \\[0.5em]
		&  = \, \sum_{z \in \Z} [z] \log \left(  \dfrac{[z]}{[z]^\text{eq}} 	\right)  - \left( [z] - [z]^\text{eq} \right) \\[0.5em]
		& = \, \mathcal{L} \left( \z \Vert \z^\text{eq} \right) \geq 0,
	\end{aligned}
\end{equation}
where we recognized the relative entropy between two non-normalized distribution $\mathcal{L} \left( \mathbf{a} \Vert \boldsymbol{b} \right) = \sum_i a_i \log (a_i/b_i) - (a_i - b_i)$.
The latter is positive and vanishes if, and only if, the two distributions are the same \cite{Cover2005}. 
Hence, the equilibrium state is the global minimum of $\G$, and the convexity of $\G$ derives from the convexity of the relative entropy.

\subsubsection{Hill-Schnakenberg decomposition}

By injecting the decomposition of the elementary fluxes in the EPR, $T \dot{\Sigma}_\R = - \F \cdot \Nabla_\R \cdot \j$, we recover the Hill-Schnakenberg decomposition of the EPR \cite{Hill1983, Schnakenberg1976}:
\begin{equation}
	\label{eq:EPR_decomposition}
	\begin{aligned}
	T \dot{\Sigma}_\R 
	& = 
	- \F_\E \cdot \Nabla \cdot \mathbb{C} \cdot \boldsymbol{J}
	- \F \cdot \Nabla_\R \cdot \invS \cdot \boldsymbol{\J} \\[.5em]
	& =
	\sum_{\varepsilon} \mathcal{A}[\boldsymbol{c}_\varepsilon] \ J^\varepsilon 
	+ 
	\sum_{z \in \Z} \mathcal{A}[\boldsymbol{g}_{z}] \ \J^{z} \\[.5em]
	& = T \dot{\Sigma}_\text{cyc} ~ + ~ T \dot{\Sigma}_\text{prod}
	\end{aligned}
\end{equation}
where, as before, $\mathcal{A}[\boldsymbol{g}_{z}]$ denotes the overall affinity along the elementary mode $\boldsymbol{g}_{z}$, and $\mathcal{A}[\boldsymbol{c}_\varepsilon]$ the overall affinity along the cycle $\boldsymbol{c}_\varepsilon$ (which vanishes if $\boldsymbol{c}_\varepsilon$ is also cycle for the external species). 
Importantly, the dissipation along the cycles of the sub-network, $T \dot{\Sigma}_\text{cyc}$, corresponds to $\dot{W}_\text{cyc}$ in Eq.~\eqref{eq:w_cyc}. 
Subsequently, the power maintaining the emergent/futile cycles of the sub-network ends up dissipated. 
On the other hand, the dissipation along the productive modes, $T \dot{\Sigma}_\text{prod}$, differs from $\dot{W}_\text{prod}$, allowing for work extraction and free-energy change in the sub-network.
Finally, at steady-state,
\begin{equation}
	T \overline{\dot{\Sigma}}_\R = 
	\sum_{\varepsilon} \mathcal{A}[\boldsymbol{c}_\varepsilon] \ \overline{J}{}^{\, \varepsilon} 
	+ 
	\sum_{y \in \Y} \mathcal{A}[\boldsymbol{g}_{y}] \ \overline{\J}{}^{\, y}.
\end{equation}
Because at steady-state, the production of the autocatalytic species must be balanced by the external fluxes, $\overline{\J}{}^{\, y} = - I^{y}$, one recovers that once NESS is settled, all the power ends up dissipated,
\begin{equation}
	T \overline{\dot{\Sigma}}_{\R} =
	\overline{\dot{W}}_\text{cyc} + 
	\overline{\dot{W}}_\text{prod},
\end{equation} 
and the free energy is constant in time $\d_t \G = 0$.

\section*{Conclusions}

In this work, we have studied the nonequilibrium behavior of autocatalytic networks by relying on the topological features of these systems.
One of the key results of this study is that, while the condition for autocatalysis is topological, it allows for an analysis of the dynamics of autocatalytic networks. 
Hence, building on recent advancements in the stoichiometric analysis of autocatalysis, we have developed a framework to describe their dynamics without requiring steady-state conditions or specific reaction kinetics.

By exploiting the hypergraph representation of an autocatalytic networks, we derived a general decomposition of chemical fluxes that is both broad and robust, revealing fundamental insights into their operation far from equilibrium.
Furthermore, the decomposition of chemical fluxes into productive modes and cycles, based solely on the network's stoichiometric matrix, highlighted two distinct types of behavior: productive fluxes that generate autocatalysts in excess, and cyclic fluxes that preserve the internal state of the network. 
This separation, dictated purely by topology, emphasizes the powerful relationship between structure and dynamics in these systems.
In addition, the structure of these networks, particularly through the examination of conservation laws, is sufficient to infer the physical roles played by different chemical species.

A significant aspect of our analysis is the role of external species coupled with the autocatalytic networks.
Indeed, stoichiometric autocatalysis, by definition, violates mass conservation, meaning that the system requires external species, which function as fuel or waste materials, enabling the autocatalytic network to operate. 
In the context of biology, these species should be constantly provided to the autocatalytic network and, in addition, external mechanisms must replenish fuel species from the excess of waste. 
Even though maintaining these external species bring non-conservative forces, we demonstrated that these forces do not apply on the productive modes of the autocatalytic networks. 
This was absent in previous analysis of open chemical networks, in which the non-conservative forces are considered homogeneously among the reactions.  
As a result, the condition for unconditional detailed balance in autocatalytic networks is independent on the non-conservative forces brought by the external species.

The thermodynamic properties of autocatalytic networks also exhibit distinct characteristics. 
From the nonequilibrium geometry of autocatalytic networks, we derived a correction that must be applied to the semigrand Gibbs free-energy to accurately describe the thermodynamics of autocatalytic networks. 
This correction arises to take into account the underlying conservative process of external species sustaining the production of autocatalysts. 
From the EPR decomposition of the thermodynamical potential, and the decomposition of the fluxes, we were able to distinguish the cost sustaining the futile cycles of autocatalytic networks from the cost sustaining the production of autocatalysts. 
While the former ends up dissipated, the later is able to generate useful work that can be extracted or change the free-energy content in the autocatalytic network.

In summary, our results underscore the central role of topology in governing both the dynamics and thermodynamics of autocatalytic networks.
By linking conservation laws, flux decomposition, and thermodynamical potentials, we have provided a comprehensive framework for understanding these systems in out-of-equilibrium conditions, with implications for both theoretical studies and practical applications in biological and chemical contexts.
Nevertheless, in the metabolism, autocatalysis generally occurs inside a compartment. 
In that case, maintaining fluxes along the productive modes will be 
induce \emph{a priori} volume growth (or shrink). 
In this work, we have neglected this effect, assuming constant volume.
In this regard, Sughiyama \emph{et al.} have recently proposed an innovative framework addressing the chemical thermodynamics of growing system in Ref.~\cite{Sughiyama2022}. 
However, it was restricted to non-singular stoichiometric matrix $\S$, we hope that the results derived here will help to extend their results in the general case, leaving the extension of this framework for future studies.

\renewcommand{\theequation}{A\arabic{equation}}
\setcounter{equation}{0}

\renewcommand{\thesubsection}{A\arabic{subsection}}
\setcounter{subsection}{0}

\section*{\label{sec:Appendix} Appendix A: General case}

\begin{figure}[H]
	\centering
	\includegraphics{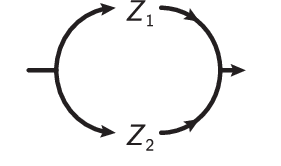}
	\caption{\label{eq:creation_annilihation}
		The non mass-like conservation laws in an autocatalytic sub-network can be traced back to the presence of \emph{creation-annihilation} motifs in the network. 
	}
\end{figure}

In the most general framework of Eq.~\eqref{eq:stoichiometric_condition}, the autocatalytic sub-network has not an empty left nullspace, $\ker ~ \S^\top \neq \left\lbrace \boldsymbol{0} \right\rbrace$ and, subsequently, it conservation laws. 
Still, Gordan's theorem \cite{Gordan1873} implies that the stoichiometric condition Eq.~\eqref{eq:stoichiometric_condition} is equivalent to the absence of \emph{mass-like} conservation laws in the sub-network. 
In other words, although not empty, the left nullspace of $\S$ is not intersecting the positive orthant: 
\begin{equation}
	\label{eq:null_intersect_positive_orthant}
	\ker ~ \S{}^\top \cap \mathbb{R}^{| \Z |}_{>0} = \varnothing.
\end{equation}
The (non mass-like) conserved quantities in autocatalytic networks are expected to be caused by creation-annihilation motifs. 
Such a motif is represented in Fig.~\ref{eq:creation_annilihation}; it consists of a complex of autocatalytic species that is formed and destroyed. 
For example, for the motif represented in Fig.~\ref{eq:creation_annilihation},
\begin{equation*}
	\boldsymbol{\ell} = 
	\begin{pNiceArray}{w{c}{1em}w{c}{1em}w{c}{1em}w{c}{1em}}[last-row]
		\boldsymbol{0} & 1 & -1 & \boldsymbol{0}  \\
		\cdots & \sf{Z}_1 & \ \sf{Z}_2 & \cdots
	\end{pNiceArray}
\end{equation*} 
is associated with the conserved quantity $L = [\sf{Z}_1] - [\sf{Z}_2]$ which expresses that species $\sf{Z}_1$ and $\sf{Z}_2$ cannot be produced independently. 
As we did with the conservation laws of the full network, we can gather a linear basis of the left nullspace of $\S$ as the rows of a matrix $\Lbar$: 
\begin{equation}
	\Lbar \cdot \S = \boldsymbol{0}. 
\end{equation} 

\subsection{Autocatalytic potential species}

Exchanging autocatalytic species with the environment is likely to break some of the non mass-like conservation laws of the sub-network. 
For example, if the motif in Fig.~\ref{eq:creation_annilihation} is present in the sub-network, its conserved quantity is broken once species $\sf{Z}_1$ (or $\sf{Z}_2$) is exchanged. 

As a result, some of the species in $\Y$ will be "potential" species,  associated with the broken conservation laws of the sub-network. 
We denote by $\Yp \subset \Y$ the subset of the autocatalytic potential species, and we let $\Yf = \Y - \Yp$ be the subset of the remaining autocatalytic force species. 
The rows of $\Lbar$ associated with species $\Yp$, $\Lbar{}^{\, \Yp}$, are the broken conservation laws of the sub-network, and, without loss of generality, we directly work with its row-reduced form:
\begin{equation}
	\label{eq:row_reduced_broken_conservation_law}
	\Lbar{}^{\, \Yp} 
	= \left( 
	\mathbbm{1}_{|\Yp|} ~,~ -\Lbar^{\Yp}_{\Yf} ~,~ -\Lbar^{\Yp}_{\X}
	\right).
\end{equation}

\subsection{Broken conservation laws}

The broken conservation laws of the network are now obtained by considering $\L$ and $\Lbar{}^{\, \Yp}$:
\begin{equation}
	\boldsymbol{\mathfrak{L}} = \hspace*{1em}
	\begin{pNiceArray}
		{m{0.8em}m{0.8em}:m{0.8em}m{0.8em}|m{1em}m{1em}:m{1em}m{1em}:m{1em}m{1em}}[first-col]
		\Vdots[line-style={solid, <->}, shorten=2pt]_{\small \rotatebox{90}{$\Ep$} }  
		& \Block{2-2}{\L_{\Ep}} &  
		& \Block{2-2}{\L_{\Ef}} & 
		& \Block{2-2}{\L_{\Yp}} & 
		& \Block{2-2}{\L_{\Yf}} & 
		& \Block{2-2}{\L_{\X}} &  
		\\
		& & 
		& & 
		& &
		& & 
		& & \\
		\Hline
		\Vdots[line-style={solid, <->}, shorten=2pt]_{\small \rotatebox{90}{$\Yp$} }
		& \Block{2-2}{\boldsymbol{0}} &  
		& \Block{2-2}{\boldsymbol{0}} & 
		& \Block{2-2}{\mathbbm{1}_{|\Yp|}} & 
		& \Block{2-2}{-\Lbar^{\Yp}_{\Yf}} & 
		& \Block{2-2}{-\Lbar_{\X}^{\Yp}} &  
		\\
		& & 
		& & 
		& & 
		& &
		& & \\
	\end{pNiceArray}.
\end{equation}
Its restriction to the potential species, $p = {\Ep \cup \Yp}$, is
\begin{equation*}
	\boldsymbol{\mathfrak{L}}_p = 
	\begin{pNiceArray}{m{1em}m{1em}|m{1.2em}m{1.2em}}
		\Block{2-2}{\L_{\Ep}} &  & \Block{2-2}{\L_{\Yp}} & \\
		& & & \\
		\Hline
		\Block{2-2}{\boldsymbol{0}} &  
		& \Block{2-2}{\mathbbm{1}_{|\Yp|}} & \\
		& & & \\
	\end{pNiceArray},
\end{equation*} 
which is a non-singular matrix:
\begin{equation*}
	{\boldsymbol{\mathfrak{L}}_p}^{-1}  =
	\begin{pNiceArray}{m{1.5em}m{1.5em}|m{1.5em}m{1.5em}}
		\Block{2-2}{\left( \L_{\Ep}\right)^{-1} } &  & \Block{2-2}{- \M_{\Yp}}& \\
		& & & \\
		\Hline
		\Block{2-2}{\boldsymbol{0}} &  & \Block{2-2}{\mathbbm{1}_{|\Yp|}} & \\
		& & & \\
	\end{pNiceArray}. 
\end{equation*}
Hence, the moiety matrix associated with all the potential species reads
\begin{equation}
	\label{eq:full_moiety_matrix}
	\begin{gathered}
		\boldsymbol{\mathfrak{M}}  = {\boldsymbol{\mathfrak{L}}_p}^{-1} \cdot \boldsymbol{\mathfrak{L}} \\[1em]
		 = \hspace{1.2em}
		\begin{pNiceArray}
			{m{.8em}m{.8em}:m{.8em}m{.8em}|m{1em}m{1em}:m{1em}m{1em}:m{1em}m{1em}}[first-col]
			\Vdots[line-style={solid, <->}, shorten=2pt]_{\small \rotatebox{90}{$\Ep$} }  
			& \Block{2-2}{\mathbbm{1}_{|\Ep|}} &  
			& \Block{2-2}{\M_{\Ef}} & 
			& \Block{2-2}{\boldsymbol{0}} & 
			& \Block{2-2}{\widehat{\M}_{\Yf}} & 
			& \Block{2-2}{\widehat{\M}_{\X}} & 
			\\
			& & 
			& & 
			& & 
			& &
			& & \\
			\Hline
			\Vdots[line-style={solid, <->}, shorten=2pt]_{\small \rotatebox{90}{$\Yp$} }
			& \Block{2-2}{\boldsymbol{0}} &  
			& \Block{2-2}{\boldsymbol{0}} & 
			& \Block{2-2}{\mathbbm{1}_{|\Yp|}} & 
			& \Block{2-2}{-\Lbar^{\Yp}_{\Yf}} & 
			& \Block{2-2}{-\Lbar_{\X}^{\Yp}} & 
			\\
			& & 
			& & 
			& & 
			& &
			& & \\
		\end{pNiceArray},
	\end{gathered}
\end{equation}
where
\begin{gather*}
	\widehat{\M}_{\Yf} =
	\M_{\Yf} + \M_{\Yp} \cdot \Lbar^{\Yp}_{\Yf}, \\[.5em]
	\widehat{\M}_\X =
	\M_\X + \M_{\Yp} \cdot \Lbar^{\Yp}_\X. 
\end{gather*}

\subsection{Elementary modes of production}

The presence of conservation laws in the sub-network prevents the definition of elementary modes for \emph{all} the autocatalytic species. 
Specifically, we can pair the conservation laws of the sub-network to a subset $\Zp \subset \Z$ of the autocatalytic species. 
Because they are associated with the broken conservation laws, the autocatalytic potential species necessarily belong to this subset, $\Yp \subset \Zp$; yet, in addition, the sub-network might still have unbroken conservation laws. 
These are related to the autocatalytic species that are not exchanged with the environment, namely the $\X$ species.  
As a result, we can define the subset of species $\Xp \subset \X$ that are associated with the unbroken conservation laws of the sub-network. 
Setting a new non-vanishing external flux on species $x_p \in \Xp$ breaks its associated conservation law, resulting in a new broken conservation law.

In the end, one has $\Zp = \Yp \cup \Xp$ while the remaining autocatalytic species are $\Zf \equiv \Z - \Zp$.
Importantly, the autocatalytic force species verify $\Yf \subset \Zf$, and finally, $\Xf \equiv \Zf - \Yf \subset \X$. 
Consequently, the stoichiometric matrix of the sub-network has the following block decomposition,
\begin{equation}
	\label{eq:block_decomposition_S}
	\S = \hspace{1em}
	\quad \begin{pNiceArray}{m{1em}m{1em}}[first-col]
		\Vdots[line-style={solid, <->}, shorten=2pt]_{\small \rotatebox{90}{$\Z_p$}} 
		& \Block{2-2}{\T} & \\
		&  & \\
		\hdottedline
		\Vdots[line-style={solid, <->}, shorten=2pt]_{\small \rotatebox{90}{$\Z_f$}}
		& \Block{2-2}{\Sbar} &  \\
		& & \\
	\end{pNiceArray}, 
	\vspace*{0.5em}
\end{equation}
in which $\Sbar$ is deprived of conservation law: $\mathrm{ker} ~\Sbar^\top = \left\lbrace \boldsymbol{0} \right\rbrace$. 
On the other hand, the conservation laws of $\S$ are associated with species $\Zp$,
\begin{equation*}
	\Lbar = 
	\left(\mathbbm{1}_{|\Zp|} ~,~  -\Lbar_{\Zf} \right). 
\end{equation*}	 
Crucially, the condition in Eq.~\eqref{eq:null_intersect_positive_orthant} implies that each column of $ \Lbar_{\Z_f} $ has at least one strictly positive entry: 
\begin{equation}
	\label{SM-eq:no_mass_like}
	\forall z_p \in \Z_p, ~~ \exists z_f \in \Z_f 
	~~\text{such that}~~ 
	\left( \Lbar_{\Z_f} \right)^{z_p}_{z_f} > 0. 
\end{equation}
As a result, the stoichiometry of species $\Zp$, $\T$, is directly related to $\Sbar$:
\begin{equation}
	\label{SM-eq:T_wrt_S}
	\T = \Lbar_{\Z_f} \cdot \Sbar.
\end{equation}
Finally, we obtain elementary modes of the species in $\Zf$ by taking the right-inverse of $\Sbar$, 
\begin{equation}
	\Sbar \cdot \invS = \mathbbm{1}_{|\Zf|}. 
\end{equation}
In addition, using Eq.~\eqref{SM-eq:T_wrt_S}, the change in species $\Zp$ along the productive modes is 
\begin{equation}
	\T \cdot \mathbb{G} = \Lbar_{ \Zf }. 
\end{equation}
Therefore, Eq.~\eqref{SM-eq:no_mass_like} insures that, for all $z_p \in \Zp$ and for every choice of right-inverse $\mathbb{G}$, there exists a species $z_f \in \Zf$ whose elementary mode, $\boldsymbol{g}_{ z_f }$, is also producing species $z_p$: 
\begin{equation}
	\left( \T \cdot \boldsymbol{g}_{ z_f } \right)^{z_p} > 0.
\end{equation}
Hence, when the most general criterion in Eq.~\eqref{eq:stoichiometric_condition} applies but not its stricter counterpart Eq.~\eqref{eq:simpler_condtion}, we can produce only a subset $\Zf$ of the autocatalytic species independently, while the remaining species $\Zp$ bear the conservation laws and are subjected to the production of a species in $\Zf$. 
When $\Zf = \Z$ and $\Zp = \varnothing$, the sub-network has no conservation and the stricter condition used in the main text holds. 

\vspace{.5\baselineskip}
\paragraph*{\textbf{Remark}}
At this point, the notations $\Xp$ and $\Xf$ are used for convenience as the species in these subsets are not exchanged with the environment, and, hence, are neither "potential" nor "force".
Nevertheless, as shown in Appendix B (\emph{cf.} the remark), at equilibrium the species in $\Xp$ (resp. $\Xf$) behave as if they were potential (resp. force) species.

\subsection{Decomposition of the fluxes}

The decomposition of the stoichiometric matrix in Eq.~\eqref{eq:block_decomposition_S} impose that
\begin{equation}
	\ker~ \S = \ker~ \Sbar, 
\end{equation}
hence, the analysis of the cycles of the sub-network made in Section~\ref{sec:cycles} of the main text is unaffected by the presence of conservation laws in the sub-network and its results apply straightforwardly. 
In particular, when $\invS^{\R_2} = \boldsymbol{0}$, the columns of $\left(\invS, \ \mathbb{C} \right)$ define a linear basis of the chemical fluxes, and the decomposition in Eq.~\eqref{eq:decomposition_fluxes} applies. 
Nevertheless, in that case, the fluxes along the productive modes is defined only on the subset of species $\Zf$ (instead of $\Z$): 
\begin{equation}
	\j = 
	\sum_{\varepsilon} J^\varepsilon (t) \ \boldsymbol{c}_\varepsilon
	+
	\sum_{z_f \in \Zf} \J^{z_f} (t) \ \boldsymbol{g}_{z_f}. 
\end{equation} 
When the sub-network approaches steady-state:
\begin{gather*}
	\forall x_f \in \Xf, ~~ \J^{x_f} (t) \longrightarrow 0, \\[.5em]
	\forall y_f \in \Yf, ~~ \J^{y_f} (t) \longrightarrow \overline{\J}{}^{\ y_f} = - I^{y_f}. 
\end{gather*} 
Such that the decomposition of the steady-fluxes becomes: 
\begin{equation}
	\label{eq:ss_fluxes_}
	\begin{gathered}
		\overline{\j} = \mathbb{C} \cdot \overline{\boldsymbol{J}} + \invS_{\Yf} \cdot \overline{\boldsymbol{\J}}{}^{\ \Yf} 
	\end{gathered}
\end{equation}

\renewcommand{\theequation}{B\arabic{equation}}
\setcounter{equation}{0}

\renewcommand{\thesubsection}{B\arabic{subsection}}
\setcounter{subsection}{0}

\section*{Appendix B: Thermodynamics of the sub-network in the general case}

\subsection{Thermodynamical potential}

In the general case, the moieties exchanged with environment are
\begin{equation}
	\label{eq:moieties_general}
	\boldsymbol{m} (\e, \z) =  \boldsymbol{\mathfrak{M}}_\E \cdot [\e] + \boldsymbol{\mathfrak{M}}_\Z \cdot [\Z].
\end{equation}
In addition, the fundamental non-conservative forces are 
\begin{equation}
	\label{eq:forces_general}
	\F = \Mu - \Mu_p \cdot \boldsymbol{\mathfrak{M}}
\end{equation}
where $\Mu_p  \equiv ( \Mu_{\Ep}, \ \Mu_{\Yp}  )$ represents the chemical potentials of all the potential species $p = \Ep \cup \Yp$.
As before, these forces vanish for the potential species: $\F_{\Ep} = \boldsymbol{0}$ and $\F_{\Yp} = \boldsymbol{0}$. 
Finally, the thermodynamical potential is given by Eq.~\eqref{eq:G_def},
\begin{equation*}
	\G (\z,  \e) = G (\z, \e) - \Mu_p \cdot \boldsymbol{m}  + \F_\E \cdot \Delta \e (\z_f).
\end{equation*}
Note that, now, the conservative influx of external species, $\Delta \e$, depends only on the species in $\Zf$, 
\begin{equation*}
	\Delta \e (\z_f) = \NablaER \cdot \invS  \cdot \Delta \z_f. 
\end{equation*}

\subsection{Autonomous sub-networks}

When the driving work rate vanishes, $\dot{W}_\text{driv} = 0$, the EPR decomposition is similar to Eq.~\eqref{eq:second_law_autonomous}; yet, in the presence of conservation laws in the sub-network, the work rate along the productive modes is now expressed only on the subset of the autocatalytic force species: 
\begin{equation}
	\dot{W}_\text{prod} = - \sum_{y_f \in \Yf} \Delta G [\boldsymbol{g}_{y_f}] \ I^{y_f}. 
\end{equation}
As a result, when 
\begin{align*}
	\widetilde{\Nabla}{}^{\, \E}_{\R_2} = \boldsymbol{0}
	& & \text{and} & & 
	\Yf = \varnothing,
\end{align*} 
$\G$ relaxes to its minimum.
Furthermore, when $\Yf = \varnothing$, the optimum of $\G$ with respect to the autocatalytic species that are not externally controlled, \emph{i.e.} $\X = \Xp \cup \Xf = \Xp \cup \Zf$, is
\begin{equation}
	\label{eq:equilibrium_from_G_}
	\begin{gathered}
		\Mu_{\Xf}^\text{eq} = 
		- \left( 
		\Mu_{\E} \cdot \Nabla^{\E}_{\R_1} + (\Mu_{\Yp}, \, \Mu_{\Xp}^\text{eq}) \cdot \T
		\right) \cdot \left( \S_{\R_1} \right)^{-1}, \\[.8em]
		\Mu_{\Xp}^\text{eq} = \Mu_{\Ep} \cdot \M_{\Xp},
	\end{gathered}
\end{equation}
which can also be obtained by balancing the chemical potentials. 
As before, the latter is indeed global minimum of $\G$:
\begin{align*}
	\Delta \G & = \G(\x, \, \y_p, \, \e) - \G(\x^\text{eq}, \, \y_p, \, \e) 
	 = \mathcal{L} \left( \x \Vert \x^\text{eq} \right) \geq 0.
\end{align*}

\vspace{.5\baselineskip}
\paragraph*{\textbf{Remark}}
Plugging the equilibrium state in the non-conservative forces yields: \\
\begin{align}
	\F^{\ \text{eq}}_{\Xp} = \boldsymbol{0},
	& \quad &
	\F_{\Xf}^{\ \text{eq}} = \F_\E \cdot \Nabla^\E_{\R_2} \cdot \left( \S_{\R_1} \right)^{-1}.
\end{align}
Hence, at equilibrium species $\Xp$ (resp. $\Xf$) behave as if they were potential (resp. force) species, having vanishing (resp. non-vanishing) non-conservative forces.

\renewcommand{\theequation}{C\arabic{equation}}
\setcounter{equation}{0}

\section*{Appendix C: Proof of Eq.~(42)}

Setting a non-vanishing external influx on all the external species the number of externally controlled species verify $\E$ verifies \cite{Rao2016}:
\begin{multline}
	|\E| = 
	\#\text{Emergent cycles in }\Nabla^\Z \\[.5em]
	+ \#\text{Broken conservation laws}.
\end{multline}
By definition, the last term is $|\Ep|$ hence, we are left with the number of emergent cycles of 
\begin{equation}
	\Nabla^\Z = \quad 
	\begin{pNiceArray}{m{1em}m{1em}|m{1em}m{1em}}[last-row]
		\Block{3-2}{\S} & & \Block{3-2}{ \boldsymbol{0} } & \\[-0.5em]
		&  &  &   \\
		& & & \\
		\Hdotsfor{2}[line-style={solid, <->}, shorten=1pt]_{ \R }   & \Hdotsfor{2}[line-style={solid, <->}, shorten=1pt]_{ \nR }  
	\end{pNiceArray}.
\end{equation}
These are the cycles of $\Nabla^\Z$ that are not also cycles of $\Nabla$.
Clearly, performing once any of the additional reactions defines an emergent cycle but, in addition, the cycles of $\S$ that are not cycles of the full network also suit.
There is $| \mathrm{ker}~\S| - |\mathrm{ker}~\widetilde{\Nabla}{}^{\E}_{\R_2} |$ of such cycles, thus 
\begin{equation}
	| \Ef | = | \mathrm{ker}~\S| - |\mathrm{ker}~\widetilde{\Nabla}{}^{\E}_{\R_2} | + |\nR|.
\end{equation}
Noticing that, $| \mathrm{ker}~\S| = |\R_2|$ yields Eq.~\eqref{eq:cardinal_Ef}. 

\vspace{.1cm}

\section*{Acknowledgment}

The author acknowledges fruitful discussions with Paul Raux, Alex Blokhuis and David Lacoste and further thanks Yann Sakref and Luis Dinis for their suggestions on an earlier draft. 


\begin{thebibliography}{51}%
	\makeatletter
	\providecommand \@ifxundefined [1]{%
		\@ifx{#1\undefined}
	}%
	\providecommand \@ifnum [1]{%
		\ifnum #1\expandafter \@firstoftwo
		\else \expandafter \@secondoftwo
		\fi
	}%
	\providecommand \@ifx [1]{%
		\ifx #1\expandafter \@firstoftwo
		\else \expandafter \@secondoftwo
		\fi
	}%
	\providecommand \natexlab [1]{#1}%
	\providecommand \enquote  [1]{``#1''}%
	\providecommand \bibnamefont  [1]{#1}%
	\providecommand \bibfnamefont [1]{#1}%
	\providecommand \citenamefont [1]{#1}%
	\providecommand \href@noop [0]{\@secondoftwo}%
	\providecommand \href [0]{\begingroup \@sanitize@url \@href}%
	\providecommand \@href[1]{\@@startlink{#1}\@@href}%
	\providecommand \@@href[1]{\endgroup#1\@@endlink}%
	\providecommand \@sanitize@url [0]{\catcode `\\12\catcode `\$12\catcode
		`\&12\catcode `\#12\catcode `\^12\catcode `\_12\catcode `\%12\relax}%
	\providecommand \@@startlink[1]{}%
	\providecommand \@@endlink[0]{}%
	\providecommand \url  [0]{\begingroup\@sanitize@url \@url }%
	\providecommand \@url [1]{\endgroup\@href {#1}{\urlprefix }}%
	\providecommand \urlprefix  [0]{URL }%
	\providecommand \Eprint [0]{\href }%
	\providecommand \doibase [0]{https://doi.org/}%
	\providecommand \selectlanguage [0]{\@gobble}%
	\providecommand \bibinfo  [0]{\@secondoftwo}%
	\providecommand \bibfield  [0]{\@secondoftwo}%
	\providecommand \translation [1]{[#1]}%
	\providecommand \BibitemOpen [0]{}%
	\providecommand \bibitemStop [0]{}%
	\providecommand \bibitemNoStop [0]{.\EOS\space}%
	\providecommand \EOS [0]{\spacefactor3000\relax}%
	\providecommand \BibitemShut  [1]{\csname bibitem#1\endcsname}%
	\let\auto@bib@innerbib\@empty
	\bibitem [{\citenamefont {Schuster}(2019)}]{Schuster2019}%
	\BibitemOpen
	\bibfield  {author} {\bibinfo {author} {\bibfnamefont {P.}~\bibnamefont
			{Schuster}},\ }\href {https://doi.org/10.1007/s00706-019-02437-z} {\bibfield
		{journal} {\bibinfo  {journal} {Monatshefte für Chemie - Chemical Monthly}\
		}\textbf {\bibinfo {volume} {150}},\ \bibinfo {pages} {763} (\bibinfo {year}
		{2019})}\BibitemShut {NoStop}%
	\bibitem [{\citenamefont {Roy}\ \emph {et~al.}(2021)\citenamefont {Roy},
		\citenamefont {Goberman},\ and\ \citenamefont {Pugatch}}]{Roy2021}%
	\BibitemOpen
	\bibfield  {author} {\bibinfo {author} {\bibfnamefont {A.}~\bibnamefont
			{Roy}}, \bibinfo {author} {\bibfnamefont {D.}~\bibnamefont {Goberman}},\ and\
		\bibinfo {author} {\bibfnamefont {R.}~\bibnamefont {Pugatch}},\ }\href
	{https://pnas.org/doi/full/10.1073/pnas.2107829118} {\bibfield  {journal}
		{\bibinfo  {journal} {Proceedings of the National Academy of Sciences}\
		}\textbf {\bibinfo {volume} {118}} (\bibinfo {year} {2021})}\BibitemShut
	{NoStop}%
	\bibitem [{\citenamefont {Lin}\ \emph {et~al.}(2020)\citenamefont {Lin},
		\citenamefont {Kussell}, \citenamefont {Young},\ and\ \citenamefont
		{Jacobs-Wagner}}]{Lin2020}%
	\BibitemOpen
	\bibfield  {author} {\bibinfo {author} {\bibfnamefont {W.-H.}\ \bibnamefont
			{Lin}}, \bibinfo {author} {\bibfnamefont {E.}~\bibnamefont {Kussell}},
		\bibinfo {author} {\bibfnamefont {L.-S.}\ \bibnamefont {Young}},\ and\
		\bibinfo {author} {\bibfnamefont {C.}~\bibnamefont {Jacobs-Wagner}},\ }\href
	{https://doi.org/10.1073/pnas.2013061117} {\bibfield  {journal} {\bibinfo
			{journal} {Proceedings of the National Academy of Sciences}\ }\textbf
		{\bibinfo {volume} {117}},\ \bibinfo {pages} {27795} (\bibinfo {year}
		{2020})}\BibitemShut {NoStop}%
	\bibitem [{\citenamefont {Sakref}\ and\ \citenamefont
		{Rivoire}(2024)}]{Sakref2024}%
	\BibitemOpen
	\bibfield  {author} {\bibinfo {author} {\bibfnamefont {Y.}~\bibnamefont
			{Sakref}}\ and\ \bibinfo {author} {\bibfnamefont {O.}~\bibnamefont
			{Rivoire}},\ }\href {https://doi.org/10.1016/j.jtbi.2023.111714} {\bibfield
		{journal} {\bibinfo  {journal} {Journal of Theoretical Biology}\ }\textbf
		{\bibinfo {volume} {579}},\ \bibinfo {pages} {111714} (\bibinfo {year}
		{2024})}\BibitemShut {NoStop}%
	\bibitem [{\citenamefont {Ameta}\ \emph {et~al.}(2021)\citenamefont {Ameta},
		\citenamefont {Matsubara}, \citenamefont {Chakraborty}, \citenamefont
		{Krishna},\ and\ \citenamefont {Thutupalli}}]{Ameta2021}%
	\BibitemOpen
	\bibfield  {author} {\bibinfo {author} {\bibfnamefont {S.}~\bibnamefont
			{Ameta}}, \bibinfo {author} {\bibfnamefont {Y.~J.}\ \bibnamefont
			{Matsubara}}, \bibinfo {author} {\bibfnamefont {N.}~\bibnamefont
			{Chakraborty}}, \bibinfo {author} {\bibfnamefont {S.}~\bibnamefont
			{Krishna}},\ and\ \bibinfo {author} {\bibfnamefont {S.}~\bibnamefont
			{Thutupalli}},\ }\href {https://doi.org/10.3390/life11040308} {\bibfield
		{journal} {\bibinfo  {journal} {Life}\ }\textbf {\bibinfo {volume} {11}},\
		\bibinfo {pages} {308} (\bibinfo {year} {2021})}\BibitemShut {NoStop}%
	\bibitem [{\citenamefont {Peng}\ \emph {et~al.}(2022)\citenamefont {Peng},
		\citenamefont {Linderoth},\ and\ \citenamefont {Baum}}]{Peng2022}%
	\BibitemOpen
	\bibfield  {author} {\bibinfo {author} {\bibfnamefont {Z.}~\bibnamefont
			{Peng}}, \bibinfo {author} {\bibfnamefont {J.}~\bibnamefont {Linderoth}},\
		and\ \bibinfo {author} {\bibfnamefont {D.~A.}\ \bibnamefont {Baum}},\ }\href
	{https://doi.org/10.1371/journal.pcbi.1010498} {\bibfield  {journal}
		{\bibinfo  {journal} {PLOS Computational Biology}\ }\textbf {\bibinfo
			{volume} {18}},\ \bibinfo {pages} {e1010498} (\bibinfo {year}
		{2022})}\BibitemShut {NoStop}%
	\bibitem [{\citenamefont {Hordijk}\ and\ \citenamefont
		{Steel}(2014)}]{Hordijk2014}%
	\BibitemOpen
	\bibfield  {author} {\bibinfo {author} {\bibfnamefont {W.}~\bibnamefont
			{Hordijk}}\ and\ \bibinfo {author} {\bibfnamefont {M.}~\bibnamefont
			{Steel}},\ }\href {https://doi.org/10.1007/s11084-014-9374-5} {\bibfield
		{journal} {\bibinfo  {journal} {Origins of Life and Evolution of Biospheres}\
		}\textbf {\bibinfo {volume} {44}},\ \bibinfo {pages} {111} (\bibinfo {year}
		{2014})}\BibitemShut {NoStop}%
	\bibitem [{\citenamefont {Vincent}\ \emph {et~al.}(2021)\citenamefont
		{Vincent}, \citenamefont {Colón-Santos}, \citenamefont {Cleaves},
		\citenamefont {Baum},\ and\ \citenamefont {Maurer}}]{Vincent2021}%
	\BibitemOpen
	\bibfield  {author} {\bibinfo {author} {\bibfnamefont {L.}~\bibnamefont
			{Vincent}}, \bibinfo {author} {\bibfnamefont {S.}~\bibnamefont
			{Colón-Santos}}, \bibinfo {author} {\bibfnamefont {H.~J.}\ \bibnamefont
			{Cleaves}}, \bibinfo {author} {\bibfnamefont {D.~A.}\ \bibnamefont {Baum}},\
		and\ \bibinfo {author} {\bibfnamefont {S.~E.}\ \bibnamefont {Maurer}},\
	}\href {https://doi.org/10.3390/life11111221} {\bibfield  {journal} {\bibinfo
			{journal} {Life}\ }\textbf {\bibinfo {volume} {11}},\ \bibinfo {pages}
		{1221} (\bibinfo {year} {2021})}\BibitemShut {NoStop}%
	\bibitem [{\citenamefont {Peng}\ \emph {et~al.}(2020)\citenamefont {Peng},
		\citenamefont {Plum}, \citenamefont {Gagrani},\ and\ \citenamefont
		{Baum}}]{Peng2020}%
	\BibitemOpen
	\bibfield  {author} {\bibinfo {author} {\bibfnamefont {Z.}~\bibnamefont
			{Peng}}, \bibinfo {author} {\bibfnamefont {A.~M.}\ \bibnamefont {Plum}},
		\bibinfo {author} {\bibfnamefont {P.}~\bibnamefont {Gagrani}},\ and\ \bibinfo
		{author} {\bibfnamefont {D.~A.}\ \bibnamefont {Baum}},\ }\href
	{https://doi.org/10.1016/j.jtbi.2020.110451} {\bibfield  {journal} {\bibinfo
			{journal} {Journal of Theoretical Biology}\ }\textbf {\bibinfo {volume}
			{507}},\ \bibinfo {pages} {110451} (\bibinfo {year} {2020})}\BibitemShut
	{NoStop}%
	\bibitem [{\citenamefont {Xavier}\ \emph {et~al.}(2020)\citenamefont {Xavier},
		\citenamefont {Hordijk}, \citenamefont {Kauffman}, \citenamefont {Steel},\
		and\ \citenamefont {Martin}}]{Xavier2020}%
	\BibitemOpen
	\bibfield  {author} {\bibinfo {author} {\bibfnamefont {J.~C.}\ \bibnamefont
			{Xavier}}, \bibinfo {author} {\bibfnamefont {W.}~\bibnamefont {Hordijk}},
		\bibinfo {author} {\bibfnamefont {S.}~\bibnamefont {Kauffman}}, \bibinfo
		{author} {\bibfnamefont {M.}~\bibnamefont {Steel}},\ and\ \bibinfo {author}
		{\bibfnamefont {W.~F.}\ \bibnamefont {Martin}},\ }\href
	{https://doi.org/10.1098/rspb.2019.2377} {\bibfield  {journal} {\bibinfo
			{journal} {Proceedings of the Royal Society B: Biological Sciences}\ }\textbf
		{\bibinfo {volume} {287}},\ \bibinfo {pages} {20192377} (\bibinfo {year}
		{2020})}\BibitemShut {NoStop}%
	\bibitem [{\citenamefont {Higgs}\ and\ \citenamefont
		{Lehman}(2015)}]{Higgs2015}%
	\BibitemOpen
	\bibfield  {author} {\bibinfo {author} {\bibfnamefont {P.~G.}\ \bibnamefont
			{Higgs}}\ and\ \bibinfo {author} {\bibfnamefont {N.}~\bibnamefont {Lehman}},\
	}\href {https://doi.org/10.1038/nrg3841} {\bibfield  {journal} {\bibinfo
			{journal} {Nature Reviews Genetics}\ }\textbf {\bibinfo {volume} {16}},\
		\bibinfo {pages} {7} (\bibinfo {year} {2015})}\BibitemShut {NoStop}%
	\bibitem [{\citenamefont {Bernhardt}(2012)}]{Bernhardt2012}%
	\BibitemOpen
	\bibfield  {author} {\bibinfo {author} {\bibfnamefont {H.~S.}\ \bibnamefont
			{Bernhardt}},\ }\href {https://doi.org/10.1186/1745-6150-7-23} {\bibfield
		{journal} {\bibinfo  {journal} {Biology Direct}\ }\textbf {\bibinfo {volume}
			{7}},\ \bibinfo {pages} {23} (\bibinfo {year} {2012})}\BibitemShut {NoStop}%
	\bibitem [{\citenamefont {Pavlinova}\ \emph {et~al.}(2023)\citenamefont
		{Pavlinova}, \citenamefont {Lambert}, \citenamefont {Malaterre},\ and\
		\citenamefont {Nghe}}]{Pavlinova2023}%
	\BibitemOpen
	\bibfield  {author} {\bibinfo {author} {\bibfnamefont {P.}~\bibnamefont
			{Pavlinova}}, \bibinfo {author} {\bibfnamefont {C.~N.}\ \bibnamefont
			{Lambert}}, \bibinfo {author} {\bibfnamefont {C.}~\bibnamefont {Malaterre}},\
		and\ \bibinfo {author} {\bibfnamefont {P.}~\bibnamefont {Nghe}},\ }\href
	{https://doi.org/10.1002/1873-3468.14507} {\bibfield  {journal} {\bibinfo
			{journal} {FEBS Letters}\ }\textbf {\bibinfo {volume} {597}},\ \bibinfo
		{pages} {344} (\bibinfo {year} {2023})}\BibitemShut {NoStop}%
	\bibitem [{\citenamefont {Kauffman}(1971)}]{Kauffman1971}%
	\BibitemOpen
	\bibfield  {author} {\bibinfo {author} {\bibfnamefont {S.~A.}\ \bibnamefont
			{Kauffman}},\ }\href {https://doi.org/10.1080/01969727108545830} {\bibfield
		{journal} {\bibinfo  {journal} {Journal of Cybernetics}\ }\textbf {\bibinfo
			{volume} {1}},\ \bibinfo {pages} {71} (\bibinfo {year} {1971})}\BibitemShut
	{NoStop}%
	\bibitem [{\citenamefont {Kauffman}(1986)}]{Kauffman1986}%
	\BibitemOpen
	\bibfield  {author} {\bibinfo {author} {\bibfnamefont {S.~A.}\ \bibnamefont
			{Kauffman}},\ }\href {https://doi.org/10.1016/S0022-5193(86)80047-9}
	{\bibfield  {journal} {\bibinfo  {journal} {Journal of Theoretical Biology}\
		}\textbf {\bibinfo {volume} {119}},\ \bibinfo {pages} {1} (\bibinfo {year}
		{1986})}\BibitemShut {NoStop}%
	\bibitem [{\citenamefont {Blokhuis}\ \emph {et~al.}(2020)\citenamefont
		{Blokhuis}, \citenamefont {Lacoste},\ and\ \citenamefont
		{Nghe}}]{Blokhuis2020}%
	\BibitemOpen
	\bibfield  {author} {\bibinfo {author} {\bibfnamefont {A.}~\bibnamefont
			{Blokhuis}}, \bibinfo {author} {\bibfnamefont {D.}~\bibnamefont {Lacoste}},\
		and\ \bibinfo {author} {\bibfnamefont {P.}~\bibnamefont {Nghe}},\ }\href
	{https://doi.org/10.1073/pnas.2013527117} {\bibfield  {journal} {\bibinfo
			{journal} {Proceedings of the National Academy of Sciences}\ }\textbf
		{\bibinfo {volume} {117}},\ \bibinfo {pages} {25230} (\bibinfo {year}
		{2020})}\BibitemShut {NoStop}%
	\bibitem [{\citenamefont {Andersen}\ \emph {et~al.}(2021)\citenamefont
		{Andersen}, \citenamefont {Flamm}, \citenamefont {Merkle},\ and\
		\citenamefont {Stadler}}]{Andersen2021}%
	\BibitemOpen
	\bibfield  {author} {\bibinfo {author} {\bibfnamefont {J.~L.}\ \bibnamefont
			{Andersen}}, \bibinfo {author} {\bibfnamefont {C.}~\bibnamefont {Flamm}},
		\bibinfo {author} {\bibfnamefont {D.}~\bibnamefont {Merkle}},\ and\ \bibinfo
		{author} {\bibfnamefont {P.~F.}\ \bibnamefont {Stadler}},\ }\href
	{https://doi.org/10.48550/arXiv.2107.03086} {\bibfield  {journal} {\bibinfo
			{journal} {arXiv [Preprint]}\ } (\bibinfo {year} {2021})}\BibitemShut
	{NoStop}%
	\bibitem [{\citenamefont {Gibbs}(1928)}]{Gibbs1928}%
	\BibitemOpen
	\bibfield  {author} {\bibinfo {author} {\bibfnamefont {J.~W.}\ \bibnamefont
			{Gibbs}},\ }\href
	{https://www.eng.uc.edu/~beaucag/Classes/AdvancedMaterialsThermodynamics/Books/Gibbs%20G.W.%20-%20Collected%20works.%20Thermodynamics.%20Volume%201-Longmans%20(1928).pdf}
	{\emph {\bibinfo {title} {The Collected Works of J. Willard Gibbs}}}\
	(\bibinfo  {publisher} {Longmans, Green and Co.},\ \bibinfo {year}
	{1928})\BibitemShut {NoStop}%
	\bibitem [{\citenamefont {de~Donder}(1927)}]{deDonder1927}%
	\BibitemOpen
	\bibfield  {author} {\bibinfo {author} {\bibfnamefont {T.}~\bibnamefont
			{de~Donder}},\ }\href {https://books.google.fr/books?id=MQ7nAAAAMAAJ} {\emph
		{\bibinfo {title} {L'affinit{\'e}}}},\ \bibinfo {series} {Memoires de la
		Classe des sciences: Collection in-8o}\ No.\ \bibinfo {number} {vol.~1}\
	(\bibinfo  {publisher} {Gauthier-Villars},\ \bibinfo {year}
	{1927})\BibitemShut {NoStop}%
	\bibitem [{\citenamefont {McQuarrie}(1967)}]{McQuarrie1967}%
	\BibitemOpen
	\bibfield  {author} {\bibinfo {author} {\bibfnamefont {D.~A.}\ \bibnamefont
			{McQuarrie}},\ }\href {https://doi.org/10.2307/3212214} {\bibfield  {journal}
		{\bibinfo  {journal} {Journal of Applied Probability}\ }\textbf {\bibinfo
			{volume} {4}},\ \bibinfo {pages} {413} (\bibinfo {year} {1967})}\BibitemShut
	{NoStop}%
	\bibitem [{\citenamefont {Gillespie}(1992)}]{Gillespie1992}%
	\BibitemOpen
	\bibfield  {author} {\bibinfo {author} {\bibfnamefont {D.~T.}\ \bibnamefont
			{Gillespie}},\ }\href {https://doi.org/10.1016/0378-4371(92)90283-V}
	{\bibfield  {journal} {\bibinfo  {journal} {Physica A: Statistical Mechanics
				and its Applications}\ }\textbf {\bibinfo {volume} {188}},\ \bibinfo {pages}
		{404} (\bibinfo {year} {1992})}\BibitemShut {NoStop}%
	\bibitem [{\citenamefont {Schnakenberg}(1976)}]{Schnakenberg1976}%
	\BibitemOpen
	\bibfield  {author} {\bibinfo {author} {\bibfnamefont {J.}~\bibnamefont
			{Schnakenberg}},\ }\href {https://doi.org/10.1103/RevModPhys.48.571}
	{\bibfield  {journal} {\bibinfo  {journal} {Reviews of Modern Physics}\
		}\textbf {\bibinfo {volume} {48}},\ \bibinfo {pages} {571} (\bibinfo {year}
		{1976})}\BibitemShut {NoStop}%
	\bibitem [{\citenamefont {Mou}\ \emph {et~al.}(1986)\citenamefont {Mou},
		\citenamefont {li~Luo},\ and\ \citenamefont {Nicolis}}]{Mou1986}%
	\BibitemOpen
	\bibfield  {author} {\bibinfo {author} {\bibfnamefont {C.~Y.}\ \bibnamefont
			{Mou}}, \bibinfo {author} {\bibfnamefont {J.}~\bibnamefont {li~Luo}},\ and\
		\bibinfo {author} {\bibfnamefont {G.}~\bibnamefont {Nicolis}},\ }\href
	{https://doi.org/10.1063/1.450623} {\bibfield  {journal} {\bibinfo  {journal}
			{The Journal of Chemical Physics}\ }\textbf {\bibinfo {volume} {84}},\
		\bibinfo {pages} {7011} (\bibinfo {year} {1986})}\BibitemShut {NoStop}%
	\bibitem [{\citenamefont {Hill}(1983)}]{Hill1983}%
	\BibitemOpen
	\bibfield  {author} {\bibinfo {author} {\bibfnamefont {T.~L.}\ \bibnamefont
			{Hill}},\ }\href {https://doi.org/10.1073/pnas.80.10.2922} {\bibfield
		{journal} {\bibinfo  {journal} {Proceedings of the National Academy of
				Sciences}\ }\textbf {\bibinfo {volume} {80}},\ \bibinfo {pages} {2922}
		(\bibinfo {year} {1983})}\BibitemShut {NoStop}%
	\bibitem [{\citenamefont {Rao}\ and\ \citenamefont {Esposito}(2016)}]{Rao2016}%
	\BibitemOpen
	\bibfield  {author} {\bibinfo {author} {\bibfnamefont {R.}~\bibnamefont
			{Rao}}\ and\ \bibinfo {author} {\bibfnamefont {M.}~\bibnamefont {Esposito}},\
	}\href {https://doi.org/10.1103/PhysRevX.6.041064} {\bibfield  {journal}
		{\bibinfo  {journal} {Physical Review X}\ }\textbf {\bibinfo {volume} {6}}
		(\bibinfo {year} {2016})}\BibitemShut {NoStop}%
	\bibitem [{\citenamefont {Ge}\ and\ \citenamefont
		{Qian}(2016{\natexlab{a}})}]{Ge2016-2}%
	\BibitemOpen
	\bibfield  {author} {\bibinfo {author} {\bibfnamefont {H.}~\bibnamefont
			{Ge}}\ and\ \bibinfo {author} {\bibfnamefont {H.}~\bibnamefont {Qian}},\
	}\href {https://doi.org/10.1016/j.chemphys.2016.03.026} {\bibfield  {journal}
		{\bibinfo  {journal} {Chemical Physics}\ }\textbf {\bibinfo {volume} {472}},\
		\bibinfo {pages} {241} (\bibinfo {year} {2016}{\natexlab{a}})}\BibitemShut
	{NoStop}%
	\bibitem [{\citenamefont {Avanzini}\ \emph {et~al.}(2021)\citenamefont
		{Avanzini}, \citenamefont {Penocchio}, \citenamefont {Falasco},\ and\
		\citenamefont {Esposito}}]{Avanzini2021}%
	\BibitemOpen
	\bibfield  {author} {\bibinfo {author} {\bibfnamefont {F.}~\bibnamefont
			{Avanzini}}, \bibinfo {author} {\bibfnamefont {E.}~\bibnamefont {Penocchio}},
		\bibinfo {author} {\bibfnamefont {G.}~\bibnamefont {Falasco}},\ and\ \bibinfo
		{author} {\bibfnamefont {M.}~\bibnamefont {Esposito}},\ }\href
	{https://pubs.aip.org/jcp/article/154/9/094114/313760/Nonequilibrium-thermodynamics-of-non-ideal}
	{\bibfield  {journal} {\bibinfo  {journal} {The Journal of Chemical Physics}\
		}\textbf {\bibinfo {volume} {154}},\ \bibinfo {pages} {94114} (\bibinfo
		{year} {2021})}\BibitemShut {NoStop}%
	\bibitem [{\citenamefont {Alberty}(2003)}]{Alberty2003}%
	\BibitemOpen
	\bibfield  {author} {\bibinfo {author} {\bibfnamefont {R.~A.}\ \bibnamefont
			{Alberty}},\ }\href {https://doi.org/10.1002/0471332607} {\emph {\bibinfo
			{title} {Thermodynamics of Biochemical Reactions}}}\ (\bibinfo  {publisher}
	{Wiley},\ \bibinfo {year} {2003})\BibitemShut {NoStop}%
	\bibitem [{\citenamefont {Qian}\ and\ \citenamefont {Beard}(2005)}]{Qian2005}%
	\BibitemOpen
	\bibfield  {author} {\bibinfo {author} {\bibfnamefont {H.}~\bibnamefont
			{Qian}}\ and\ \bibinfo {author} {\bibfnamefont {D.~A.}\ \bibnamefont
			{Beard}},\ }\href {https://doi.org/10.1016/j.bpc.2004.12.001} {\bibfield
		{journal} {\bibinfo  {journal} {Biophysical Chemistry}\ }\textbf {\bibinfo
			{volume} {114}},\ \bibinfo {pages} {213} (\bibinfo {year}
		{2005})}\BibitemShut {NoStop}%
	\bibitem [{\citenamefont {Cengio}\ \emph {et~al.}(2023)\citenamefont {Cengio},
		\citenamefont {Lecomte},\ and\ \citenamefont {Polettini}}]{Cengio2022}%
	\BibitemOpen
	\bibfield  {author} {\bibinfo {author} {\bibfnamefont {S.~D.}\ \bibnamefont
			{Cengio}}, \bibinfo {author} {\bibfnamefont {V.}~\bibnamefont {Lecomte}},\
		and\ \bibinfo {author} {\bibfnamefont {M.}~\bibnamefont {Polettini}},\ }\href
	{https://doi.org/10.1103/PhysRevX.13.021040} {\bibfield  {journal} {\bibinfo
			{journal} {Physical Review X}\ }\textbf {\bibinfo {volume} {13}},\ \bibinfo
		{pages} {021040} (\bibinfo {year} {2023})}\BibitemShut {NoStop}%
	\bibitem [{\citenamefont {Harunari}\ \emph {et~al.}(2022)\citenamefont
		{Harunari}, \citenamefont {Dutta}, \citenamefont {Polettini},\ and\
		\citenamefont {Édgar Roldán}}]{Harunari2022}%
	\BibitemOpen
	\bibfield  {author} {\bibinfo {author} {\bibfnamefont {P.~E.}\ \bibnamefont
			{Harunari}}, \bibinfo {author} {\bibfnamefont {A.}~\bibnamefont {Dutta}},
		\bibinfo {author} {\bibfnamefont {M.}~\bibnamefont {Polettini}},\ and\
		\bibinfo {author} {\bibnamefont {Édgar Roldán}},\ }\href
	{https://doi.org/10.1103/PhysRevX.12.041026} {\bibfield  {journal} {\bibinfo
			{journal} {Physical Review X}\ }\textbf {\bibinfo {volume} {12}},\ \bibinfo
		{pages} {041026} (\bibinfo {year} {2022})}\BibitemShut {NoStop}%
	\bibitem [{\citenamefont {Aslyamov}\ and\ \citenamefont
		{Esposito}(2024)}]{Aslyamov2024}%
	\BibitemOpen
	\bibfield  {author} {\bibinfo {author} {\bibfnamefont {T.}~\bibnamefont
			{Aslyamov}}\ and\ \bibinfo {author} {\bibfnamefont {M.}~\bibnamefont
			{Esposito}},\ }\href {https://doi.org/10.1103/PhysRevLett.132.037101}
	{\bibfield  {journal} {\bibinfo  {journal} {Physical Review Letters}\
		}\textbf {\bibinfo {volume} {132}},\ \bibinfo {pages} {037101} (\bibinfo
		{year} {2024})}\BibitemShut {NoStop}%
	\bibitem [{\citenamefont {Owen}\ and\ \citenamefont
		{Horowitz}(2023)}]{Owen2023}%
	\BibitemOpen
	\bibfield  {author} {\bibinfo {author} {\bibfnamefont {J.~A.}\ \bibnamefont
			{Owen}}\ and\ \bibinfo {author} {\bibfnamefont {J.~M.}\ \bibnamefont
			{Horowitz}},\ }\href {https://doi.org/10.1038/s41467-023-36705-8} {\bibfield
		{journal} {\bibinfo  {journal} {Nature communications}\ }\textbf {\bibinfo
			{volume} {14}},\ \bibinfo {pages} {1280} (\bibinfo {year}
		{2023})}\BibitemShut {NoStop}%
	\bibitem [{\citenamefont {Despons}\ \emph {et~al.}(2023)\citenamefont
		{Despons}, \citenamefont {de~Decker},\ and\ \citenamefont
		{Lacoste}}]{Despons2023}%
	\BibitemOpen
	\bibfield  {author} {\bibinfo {author} {\bibfnamefont {A.}~\bibnamefont
			{Despons}}, \bibinfo {author} {\bibfnamefont {Y.}~\bibnamefont {de~Decker}},\
		and\ \bibinfo {author} {\bibfnamefont {D.}~\bibnamefont {Lacoste}},\ }\href
	{http://arxiv.org/abs/2306.02366} {\bibfield  {journal} {\bibinfo  {journal}
			{arXiv [Preprint]}\ } (\bibinfo {year} {2023})}\BibitemShut {NoStop}%
	\bibitem [{\citenamefont {Feinberg}(2019)}]{Feinberg2019}%
	\BibitemOpen
	\bibfield  {author} {\bibinfo {author} {\bibfnamefont {M.}~\bibnamefont
			{Feinberg}},\ }\href {https://doi.org/10.1007/978-3-030-03858-8} {\emph
		{\bibinfo {title} {Foundations of Chemical Reaction Network Theory}}}\
	(\bibinfo  {publisher} {Springer International Publishing},\ \bibinfo {year}
	{2019})\BibitemShut {NoStop}%
	\bibitem [{\citenamefont {Epstein}\ and\ \citenamefont
		{Pojman}(1998)}]{Epstein1998}%
	\BibitemOpen
	\bibfield  {author} {\bibinfo {author} {\bibfnamefont {I.~R.}\ \bibnamefont
			{Epstein}}\ and\ \bibinfo {author} {\bibfnamefont {J.~A.}\ \bibnamefont
			{Pojman}},\ }\href {https://doi.org/10.1093/oso/9780195096705.001.0001}
	{\emph {\bibinfo {title} {An Introduction to Nonlinear Chemical Dynamics}}}\
	(\bibinfo  {publisher} {Oxford University Press},\ \bibinfo {year}
	{1998})\BibitemShut {NoStop}%
	\bibitem [{\citenamefont {Kurtz}(1972)}]{Kurtz1972}%
	\BibitemOpen
	\bibfield  {author} {\bibinfo {author} {\bibfnamefont {T.~G.}\ \bibnamefont
			{Kurtz}},\ }\href {https://doi.org/10.1063/1.1678692} {\bibfield  {journal}
		{\bibinfo  {journal} {The Journal of Chemical Physics}\ }\textbf {\bibinfo
			{volume} {57}},\ \bibinfo {pages} {2976} (\bibinfo {year}
		{1972})}\BibitemShut {NoStop}%
	\bibitem [{\citenamefont {Ge}\ and\ \citenamefont
		{Qian}(2016{\natexlab{b}})}]{Ge2016}%
	\BibitemOpen
	\bibfield  {author} {\bibinfo {author} {\bibfnamefont {H.}~\bibnamefont
			{Ge}}\ and\ \bibinfo {author} {\bibfnamefont {H.}~\bibnamefont {Qian}},\
	}\href {https://doi.org/10.1103/PhysRevE.94.052150} {\bibfield  {journal}
		{\bibinfo  {journal} {Physical Review E}\ }\textbf {\bibinfo {volume} {94}},\
		\bibinfo {pages} {3} (\bibinfo {year} {2016}{\natexlab{b}})}\BibitemShut
	{NoStop}%
	\bibitem [{\citenamefont {Peka\v{r}}(2005)}]{Pekar2005}%
	\BibitemOpen
	\bibfield  {author} {\bibinfo {author} {\bibfnamefont {M.}~\bibnamefont
			{Peka\v{r}}},\ }\href {https://doi.org/10.3184/007967405777874868} {\bibfield
		{journal} {\bibinfo  {journal} {Progress in Reaction Kinetics and
				Mechanism}\ }\textbf {\bibinfo {volume} {30}},\ \bibinfo {pages} {3}
		(\bibinfo {year} {2005})}\BibitemShut {NoStop}%
	\bibitem [{\citenamefont {Rao}\ and\ \citenamefont {Esposito}(2018)}]{Rao2018}%
	\BibitemOpen
	\bibfield  {author} {\bibinfo {author} {\bibfnamefont {R.}~\bibnamefont
			{Rao}}\ and\ \bibinfo {author} {\bibfnamefont {M.}~\bibnamefont {Esposito}},\
	}\href {https://doi.org/10.1063/1.5042253} {\bibfield  {journal} {\bibinfo
			{journal} {Journal of Chemical Physics}\ }\textbf {\bibinfo {volume} {149}},\
		\bibinfo {pages} {245101} (\bibinfo {year} {2018})}\BibitemShut {NoStop}%
	\bibitem [{\citenamefont {Haraldsdóttir}\ and\ \citenamefont
		{Fleming}(2016)}]{Haraldsdottir2016}%
	\BibitemOpen
	\bibfield  {author} {\bibinfo {author} {\bibfnamefont {H.~S.}\ \bibnamefont
			{Haraldsdóttir}}\ and\ \bibinfo {author} {\bibfnamefont {R.~M.~T.}\
			\bibnamefont {Fleming}},\ }\href
	{https://doi.org/10.1371/journal.pcbi.1004999} {\bibfield  {journal}
		{\bibinfo  {journal} {PLOS Computational Biology}\ }\textbf {\bibinfo
			{volume} {12}},\ \bibinfo {pages} {e1004999} (\bibinfo {year}
		{2016})}\BibitemShut {NoStop}%
	\bibitem [{\citenamefont {Avanzini}\ and\ \citenamefont
		{Esposito}(2022)}]{Avanzini2022}%
	\BibitemOpen
	\bibfield  {author} {\bibinfo {author} {\bibfnamefont {F.}~\bibnamefont
			{Avanzini}}\ and\ \bibinfo {author} {\bibfnamefont {M.}~\bibnamefont
			{Esposito}},\ }\href
	{https://pubs.aip.org/jcp/article/156/1/014116/2839684/Thermodynamics-of-concentration-vs-flux-control-in}
	{\bibfield  {journal} {\bibinfo  {journal} {The Journal of Chemical Physics}\
		}\textbf {\bibinfo {volume} {156}} (\bibinfo {year} {2022})}\BibitemShut
	{NoStop}%
	\bibitem [{\citenamefont {Nelson}\ and\ \citenamefont
		{Cox}(2024)}]{Lehninger2024}%
	\BibitemOpen
	\bibfield  {author} {\bibinfo {author} {\bibfnamefont {D.~L.}\ \bibnamefont
			{Nelson}}\ and\ \bibinfo {author} {\bibfnamefont {M.}~\bibnamefont {Cox}},\
	}\href@noop {} {\emph {\bibinfo {title} {Principles of Biochemistry}}}\
	(\bibinfo  {publisher} {W.H Freeman and Co.},\ \bibinfo {year} {2024})\
	Chap.~\bibinfo {chapter} {16}\BibitemShut {NoStop}%
	\bibitem [{\citenamefont {Polettini}\ and\ \citenamefont
		{Esposito}(2014)}]{Polettini2014}%
	\BibitemOpen
	\bibfield  {author} {\bibinfo {author} {\bibfnamefont {M.}~\bibnamefont
			{Polettini}}\ and\ \bibinfo {author} {\bibfnamefont {M.}~\bibnamefont
			{Esposito}},\ }\href
	{https://pubs.aip.org/jcp/article/141/2/024117/352438/Irreversible-thermodynamics-of-open-chemical}
	{\bibfield  {journal} {\bibinfo  {journal} {The Journal of Chemical Physics}\
		}\textbf {\bibinfo {volume} {141}} (\bibinfo {year} {2014})}\BibitemShut
	{NoStop}%
	\bibitem [{\citenamefont {Wachtel}\ \emph {et~al.}(2018)\citenamefont
		{Wachtel}, \citenamefont {Rao},\ and\ \citenamefont
		{Esposito}}]{Wachtel2018}%
	\BibitemOpen
	\bibfield  {author} {\bibinfo {author} {\bibfnamefont {A.}~\bibnamefont
			{Wachtel}}, \bibinfo {author} {\bibfnamefont {R.}~\bibnamefont {Rao}},\ and\
		\bibinfo {author} {\bibfnamefont {M.}~\bibnamefont {Esposito}},\ }\href
	{https://doi.org/10.1088/1367-2630/aab5c9} {\bibfield  {journal} {\bibinfo
			{journal} {New Journal of Physics}\ }\textbf {\bibinfo {volume} {20}},\
		\bibinfo {pages} {042002} (\bibinfo {year} {2018})}\BibitemShut {NoStop}%
	\bibitem [{\citenamefont {Sughiyama}\ \emph {et~al.}(2022)\citenamefont
		{Sughiyama}, \citenamefont {Kamimura}, \citenamefont {Loutchko},\ and\
		\citenamefont {Kobayashi}}]{Sughiyama2022}%
	\BibitemOpen
	\bibfield  {author} {\bibinfo {author} {\bibfnamefont {Y.}~\bibnamefont
			{Sughiyama}}, \bibinfo {author} {\bibfnamefont {A.}~\bibnamefont {Kamimura}},
		\bibinfo {author} {\bibfnamefont {D.}~\bibnamefont {Loutchko}},\ and\
		\bibinfo {author} {\bibfnamefont {T.~J.}\ \bibnamefont {Kobayashi}},\ }\href
	{https://doi.org/10.1103/PhysRevResearch.4.033191} {\bibfield  {journal}
		{\bibinfo  {journal} {Physical Review Research}\ }\textbf {\bibinfo {volume}
			{4}},\ \bibinfo {pages} {033191} (\bibinfo {year} {2022})}\BibitemShut
	{NoStop}%
	\bibitem [{\citenamefont {Avanzini}\ \emph {et~al.}(2023)\citenamefont
		{Avanzini}, \citenamefont {Freitas},\ and\ \citenamefont
		{Esposito}}]{Avanzini2023}%
	\BibitemOpen
	\bibfield  {author} {\bibinfo {author} {\bibfnamefont {F.}~\bibnamefont
			{Avanzini}}, \bibinfo {author} {\bibfnamefont {N.}~\bibnamefont {Freitas}},\
		and\ \bibinfo {author} {\bibfnamefont {M.}~\bibnamefont {Esposito}},\ }\href
	{https://doi.org/10.1103/PhysRevX.13.021041} {\bibfield  {journal} {\bibinfo
			{journal} {Physical Review X}\ }\textbf {\bibinfo {volume} {13}},\ \bibinfo
		{pages} {021041} (\bibinfo {year} {2023})}\BibitemShut {NoStop}%
	\bibitem [{\citenamefont {Raux}\ \emph {et~al.}(2024)\citenamefont {Raux},
		\citenamefont {Goupil},\ and\ \citenamefont {Verley}}]{Raux2024}%
	\BibitemOpen
	\bibfield  {author} {\bibinfo {author} {\bibfnamefont {P.}~\bibnamefont
			{Raux}}, \bibinfo {author} {\bibfnamefont {C.}~\bibnamefont {Goupil}},\ and\
		\bibinfo {author} {\bibfnamefont {G.}~\bibnamefont {Verley}},\ }\href
	{https://doi.org/10.1103/PhysRevE.110.014134} {\bibfield  {journal} {\bibinfo
			{journal} {Phys. Rev. E}\ }\textbf {\bibinfo {volume} {110}},\ \bibinfo
		{pages} {014134} (\bibinfo {year} {2024})}\BibitemShut {NoStop}%
	\bibitem [{\citenamefont {Wachtel}\ \emph {et~al.}(2022)\citenamefont
		{Wachtel}, \citenamefont {Rao},\ and\ \citenamefont
		{Esposito}}]{Wachtel2022}%
	\BibitemOpen
	\bibfield  {author} {\bibinfo {author} {\bibfnamefont {A.}~\bibnamefont
			{Wachtel}}, \bibinfo {author} {\bibfnamefont {R.}~\bibnamefont {Rao}},\ and\
		\bibinfo {author} {\bibfnamefont {M.}~\bibnamefont {Esposito}},\ }\href
	{https://doi.org/10.1063/5.0091035} {\bibfield  {journal} {\bibinfo
			{journal} {The Journal of Chemical Physics}\ }\textbf {\bibinfo {volume}
			{157}} (\bibinfo {year} {2022})}\BibitemShut {NoStop}%
	\bibitem [{\citenamefont {Cover}\ and\ \citenamefont
		{Thomas}(2005)}]{Cover2005}%
	\BibitemOpen
	\bibfield  {author} {\bibinfo {author} {\bibfnamefont {T.~M.}\ \bibnamefont
			{Cover}}\ and\ \bibinfo {author} {\bibfnamefont {J.~A.}\ \bibnamefont
			{Thomas}},\ }\href {https://doi.org/10.1002/047174882X} {\emph {\bibinfo
			{title} {Elements of Information Theory}}}\ (\bibinfo  {publisher} {Wiley},\
	\bibinfo {year} {2005})\BibitemShut {NoStop}%
	\bibitem [{\citenamefont {Gordan}(1873)}]{Gordan1873}%
	\BibitemOpen
	\bibfield  {author} {\bibinfo {author} {\bibfnamefont {P.}~\bibnamefont
			{Gordan}},\ }\href {https://doi.org/10.1007/BF01442864} {\bibfield  {journal}
		{\bibinfo  {journal} {Mathematische Annalen}\ }\textbf {\bibinfo {volume}
			{6}},\ \bibinfo {pages} {23} (\bibinfo {year} {1873})}\BibitemShut {NoStop}%
\end{thebibliography}

%

\end{document}